\documentclass[lettersize,journal]{IEEEtran}
\usepackage{amsmath,amsfonts}
\usepackage{algorithmic}
\usepackage{array}
\usepackage[caption=false,font=normalsize,labelfont=sf,textfont=sf]{subfig}
\usepackage{textcomp}
\usepackage{stfloats}
\usepackage{url}
\usepackage{verbatim}
\usepackage{graphicx}
\usepackage{multirow}
\usepackage{hyperref}
\usepackage{multicol}
\usepackage{float}
\usepackage{balance}
\usepackage{amsmath,amsfonts}
\usepackage{algorithmic}
\usepackage{array}
\usepackage[caption=false,font=normalsize,
   labelfont=sf,textfont=sf]{subfig}
\usepackage{textcomp}
\usepackage{stfloats}
\usepackage{url}
\usepackage{verbatim}
\usepackage{graphicx}
\usepackage{balance}
\def\BibTeX{{\rm B\kern-.05em{\sc i\kern-.025em b}\kern-.08em
    T\kern-.1667em\lower.7ex\hbox{E}\kern-.125emX}}
\usepackage{balance}

\DeclareMathOperator*{\maxa}{\,max}

\DeclareMathOperator*{\mini}{\,min}
\begin{document}

\title{Modeling Cognitive-Affective Processes with Appraisal and Reinforcement Learning}

\author{Jiayi Zhang, Joost Broekens, Jussi Jokinen}


\maketitle

\begin{abstract}
Computational models can advance affective science by shedding light onto the interplay between cognition and emotion from an information processing point of view.
We propose a computational model of emotion that integrates reinforcement learning (RL) and appraisal theory, establishing a formal relationship between reward processing, goal-directed task learning, cognitive appraisal, and emotional experiences.
The model achieves this by formalizing three evaluative checks from the component process model (CPM) in terms of temporal difference learning updates: goal relevance, goal conduciveness, and power.
The formalism is task independent and can be applied to any task that is represented as a Markov decision problem (MDP) and solved using RL.
We evaluate the model by predicting a range of human emotions based on a series of vignette studies, highlighting its potential to improve our understanding of the role of reward processing in affective experiences.  
  \end{abstract}


\section{Introduction}

\IEEEPARstart{C}{omputational} cognitive models of emotion contribute significantly to the field of affective computing \cite{elliott1992affective,gebhard2005alma,marsella2009ema}.
They formalize hypotheses linking cognitive processes to emotional responses.
This elucidates the interplay between goal-oriented behavior, cognitive processing, and emotional states.
Such understanding improves the precision of emotion prediction in affective computing, crucial for machines that adapt to their users \cite{howes2023towards}.
Emotions have an integral role in human goal-directed behavior and problem-solving, motivating actions and providing explanatory frameworks \cite{izard2009emotion, moors2019demystifying}.
They also participate in the feedback mechanisms underlying learning and adaptation, essential for effective task performance.
However, modeling emotion's role in motivation and adaptation is challenging due to the latent nature of cognitive processes.
The expansive theoretical space linking task events to emotional responses requires robust priors for tractable modeling.

In this paper, we develop a computational cognitive model that simulates emotion elicitation within goal-oriented interactive tasks.
The model addresses the interplay between cognition and emotion by integrating a component process model (CPM) of emotional appraisal with a reinforcement learning (RL) framework for goal-directed behavior.
Appraisal theory posits emotion as an evaluative cognitive process \cite{lazarus1991emotion}.
CPM offers a detailed account of this evaluation, analyzing it into cognitive checks that assess event significance and coping capacities \cite{scherer2009dynamic, sander2005systems}.
RL serves as a computational framework for decision-making in complex settings, outlining adaptive behavior through learning \cite{sutton2018reinforcement}.
Our model's key contribution is the operationalization of specific CPM checks via RL computations.
Consequently, we present emotion as inherently coupled with goal-oriented behavior, emerging from the same adaptive processes.
The paper investigates the extent to which this CPM-RL integration replicates emotional responses in interactive task environments.

Our work builds on two key theoretical insights.
First, we treat emotional appraisal as a dynamic cognitive process that assesses event characteristics to predict emotions.
Existing models like the Emotion and Adaptation (EMA) model \cite{marsella2009ema} accomplish this by computing factors such as relevance and desirability and deriving emotions from these computed patterns.
However, these models lack an autonomous agent capable of evaluating actions to optimize expected outcomes in complex interaction.
The second insight of our model fills this gap by conceptualizing emotions as a manifestation of reward processing, particularly within the computational evaluation of how an event subjectively alters situational prospects \cite{broekens2018temporal}.

\begin{figure*}[!t]
    \centering
    \includegraphics[width=17.5cm]{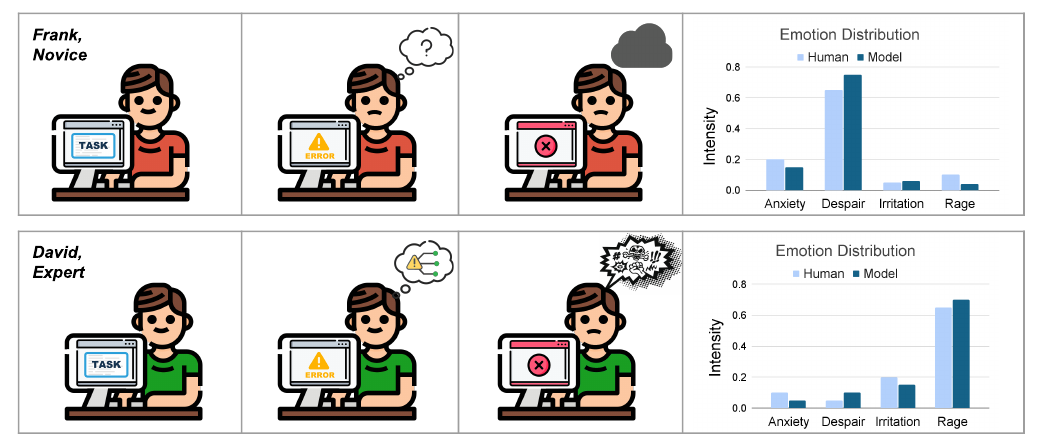}
    \caption{
      Emotional response to an event may vary based on cognitive factors.
      In the top row, Frank, an inexperienced trainee, encounters a fatal computer problem during an important work task, prompting an appraisal response and resulting in Frank feeling desperation.
      At the bottom, David, who is an expert, will have a different emotional response to the same event due to the appraisal response checking that David has some power to deal with the situation, yet the result is still obstructive to David's goals.
      Our model predicts emotional responses based on the decision process and event outcomes, considering key appraisals such as suddenness, goal relevance, conduciveness, and power. 
      The model calculates these values to generate an appraisal vector, enabling the prediction of the resulting emotion.
      The divergent emotional predictions for Frank and David primarily hinge on their contrasting levels of power. Within the RL framework, power is conceptualized as an agent's ability to choose actions that can influence its environment or outcomes, subsequently affecting its reward. 
      Model predictions are matched with human data from vignette experiments.
}
    \label{fig:fig1}
\end{figure*}

To illustrate the types of problems motivating our work and our model's solution, consider a scenario where a goal-oriented interaction is abruptly interrupted (Figure \ref{fig:fig1}).
Frank, a novice trainee, faces a challenging computer error while working on an important project.
His inexperience renders him powerless, leading to feelings of desperation.
In contrast, David, a seasoned expert, encounters the same error but reacts with anger.
He has the expertise to solve the issue but recognizes the time cost involved.
The critical difference in their emotional responses hinges on their respective levels of power over the situation.
Our model addresses this variance by computing CPM checks through RL updates.
The error serves as a negative feedback signal, which, when combined with individual assessment factors like perceived power, generates an appraisal pattern.
This pattern maps onto a range of possible emotions.
In our example, the same event elicits differing emotions: Frank's low perceived power steers him towards desperation, while David's higher level of power inclines him towards frustration and anger.


Our model makes the following contributions to the state-of-the-art in emotion modeling:
\begin{itemize}
\item Integration of the component process model (CPM) with reinforcement learning (RL), providing a computational architecture for predicting emotional responses in specific interactive tasks.
\item Introduction of a formal computational framework for four key appraisal components: suddenness, goal relevance, conduciveness, and power.
\item Empirical validation of the model's predictions through human data collected from a series of vignette experiments.
\end{itemize}

\section{Related work}
\subsection{Emotion and Cognition}


The influence of cognition on human emotions has been recognized since the 1960s, primarily through the introduction of appraisal theories highlighting the role of cognitive evaluation in emotional experiences \cite{arnold1960emotion}.
This framework has been instrumental for understanding the cognitive aspects of stress and emotion regulation \cite{lazarus1984stress}.
It asserts that emotional experiences originate from the appraisal of situational importance, a cognitive event reliant on information processing \cite{frijda1988laws}.

Appraisal theories can be expressed as models that integrate information from diverse sources like senses, memory, and reasoning, culminating in emotion as a dynamic process rather than a static state \cite{folkman1985if,scherer1982emotion,scherer2005emotions}.
Throughout this process, multiple appraisal dimensions such as goal relevance and coping abilities are assessed.
Though these models offer an abstract depiction of the appraisal information flow, neural correlates have been identified, linking brain computations to specific appraisal operations \cite{rolls2000brain,sander2018appraisal}.

Humans employ appraisal mechanisms to assess environmental stimuli and trigger emotional responses, while artificial agents use reward functions to evaluate virtual environments and determine actions \cite{sequeira2015emergence}.
This parallel between human appraisal and agent reward functions provides a lens for designing and interpreting AI behavior.
This view enhances both the understanding of agent-environment interactions and the potential for developing more human-aligned, adaptive AI systems \cite{howes2023towards}.

Understanding the psychological underpinnings of human emotion is crucial for the design of interactive systems.
Design features can elicit both positive and negative emotions, affecting user satisfaction and engagement with technology \cite{isbister2016games,norman2004emotional}.
Affective computing focuses on the recognition, interpretation, and expression of emotions in computer systems to improve user experiences \cite{picard2003affective}.
Various emotions, such as joy \cite{hassenzahl2006user}, frustration \cite{ceaparu2004determining}, pride \cite{nacke2008flow}, shame \cite{schneeberger2019can}, boredom \cite{giakoumis2010identifying}, and confusion \cite{dmello2014confusion}, arise during interactions with technology.
Their detection has become a key research area in HCI \cite{calvo2010affect}.
However, given that emotion is a mental process, its detection based solely on observable behavior is limiting.
A computational model is needed to articulate hypotheses about how latent user states, like goals and knowledge, interact with observed behavior to generate emotions.
Despite progress in understanding emotion in human-computer interactions, a computational framework linking interactive events, user cognition, and emotional outcomes remains to be developed.


The Component Process Model (CPM) by Scherer \cite{scherer2009dynamic} provides a structured approach to understanding appraisal.
It systematically dissects evaluative processes to gauge an event's personal significance \cite{sander2005systems,scherer2018studying}.
The model consists of four appraisal check classes: relevance, implications, coping potential, and normative significance, each evaluating specific facets of an event in relation to individual goals, capabilities, and societal norms.
For instance, relevance appraisals consider novelty and intrinsic pleasantness, while implications focus on goal-conduciveness.
Coping potential assesses the agent's capacity to manage the event and its potential outcomes, and whether anyone would have such capacity.
Normative significance evaluates the event's compatibility with internal and external standards, such as cultural norms.
In total, the CPM specifies 14 individual appraisal checks \cite{sander2005systems,scherer2009dynamic}.
Upon encountering a stimulus, the individual's appraisal process examines these checks, shaping the resultant emotional response.

Appraisal theory, particularly as articulated through the CPM, excels in identifying the cognitive basis of emotional experiences in interactive settings.
In contrast to basic emotion theory, which focuses on physiological patterns and corresponding basic emotions \cite{ekman1992argument}, and core affect theory, which emphasizes a two-dimensional core affect \cite{russell2003core}, appraisal theory specifies the cognitive variables and processes shaping emotional responses.
Though valuable for affective computing, especially in sensor-based emotion detection \cite{picard2003affective}, both basic emotion and core affect theories fall short in detailing the cognitive dimensions of emotion elicitation.
Appraisal theory does not confine itself to a small set of basic emotions but acknowledges that specific appraisal patterns recur frequently, thereby meriting emotion labels for easier representation and communication.
These are termed \emph{modal emotions}, which encompass not only basic emotions like joy and disgust but also other emotions distinguished by unique but frequent appraisal patterns \cite{scherer2005emotions}.


\subsection{Computational Models of Emotion} 


Affective computing has made notable strides in sensory-based emotion prediction, successfully identifying cues like facial expressions and vocal tones \cite{picard2003affective, calvo2010affect, zeng2007survey}.
However, the field still lacks reliable, general emotion sensing capabilities \cite{dmello2015review, shu2018review}.
We argue, along with others \cite{dudzik2023,dudzik2020}, that this limitation stems from an over-reliance on bodily signals, which overlooks the complex role of latent cognitive processes in shaping emotion.
A way to address this limitation is to model the latent processes that cause the emotional responses associated with the observable physical patterns in the body.

Emotions arise from a dynamic interplay between cognitive appraisals and emotional experiences, necessitating a computational architecture that captures this complexity \cite{scherer2009emotions}.
The Emotion and Adaptation (EMA) model highlights the role of appraisal processes in emotion generation \cite{marsella2009ema}.
This model has been employed across domains like virtual agents and affective computing to simulate and predict emotional responses grounded in cognitive appraisals \cite{gratch2004domain}.
The OCC model by Ortony, Clore, and Collins identifies 22 unique emotions stemming from the appraisal of three kinds of events: goal-relevant events, agent actions, and object aspects \cite{ortony1988cognitive}.
Its applicability spans AI, virtual agents, and affective computing \cite{elliott1992affective, gebhard2005alma, dias2005feeling, popescu2013gamygdala}.
Meanwhile, RL-based models use the notion of reward-processing to simulate learning and adaptation, potentially capturing the dynamic, adaptive character of human emotion and cognition \cite{moerland2017emotion,moerland2018emotion}.


Existing approaches like the OCC and EMA have demonstrated success in emotion modeling, but may lack the nuanced representation needed to capture the goal-directed nature of human interaction and its relation to emotional response \cite{moors2019demystifying}.
There is an evident gap in models that can both predict realistic behavioral trajectories in a goal-directed manner and also model the reward-based learning mechanisms that underpin human-like emotional responses \cite{moerland2018emotion}.
Integrating appraisal and RL can bridge this gap, allowing for dynamic, context-sensitive modeling of emotional responses, along with the ability to generate and justify behavioral trajectories based on goals, abilities, and the task environment.

\section{Model}
\noindent At the core of our approach is the integration of appraisal processes with RL, which emphasizes the adaptation of organisms and the associated emotional processes.
RL provides a computational framework for modeling an agent's learning and decision-making processes by updating value expectations for actions in specific situations.
These value expectations are intrinsically linked to the goals that guide an agent's behavior, and thus play a vital role in the learning process.
In contrast, appraisal theory addresses emotion from a goal-directed perspective, examining the evaluation of events and their implications, for instance, concerning personal relevance and coping potential.
Our model combines RL with appraisal, simulating how adaptation and generation of learning signals that update value expectations are connected to emotions. This fusion not only enriches our understanding of human decision-making but also grounds emotion simulation in the adaptive capabilities of AI systems.

\begin{figure}[!t]
    \centering
    \includegraphics[width=9cm]{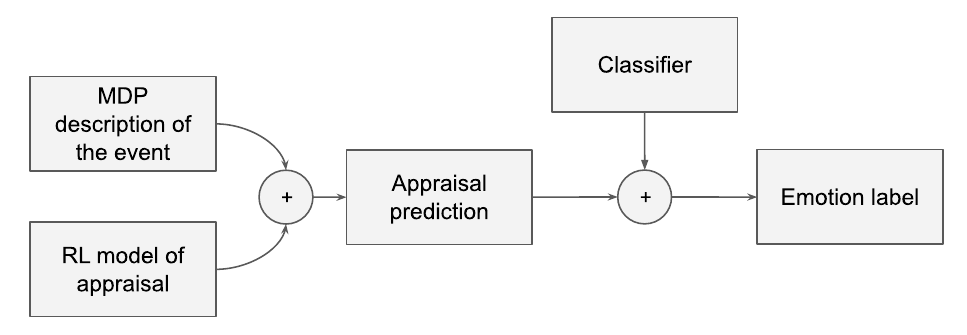}
    \caption{An overview of our computational model of emotion. A reinforcement learning agent is trained within a task environment described as a Markov decision process. Learning signals are transformed into an appraisal prediction, which can be labeled with the assistance of a pre-trained classifier.}
    \label{fig:model}
\end{figure}

Figure \ref{fig:model} presents an overview of our model.
Employing a mathematical formalism to describe task environments and model the learning of an agent within a specific environment, the model predicts appraisal ``checks'' as a result of computing learning signals.
These predictions are categorized based on established connections between appraisal patterns and emotion labels, empirically reported by Scherer and his colleagues \cite{scherer1999appraisal,scherer2001appraisal_book}.
Consequently, the model can predict the intensity of the agent's experience of particular emotions, such as joy or frustration, in response to an event during the interaction that the model simulates.


A Markov decision process (MDP) is a mathematical framework for modeling decision-making problems in stochastic environments \cite{sutton2018reinforcement}.
It is formally defined as a tuple $(S, A, T, R, \gamma)$, where $S$ denotes the set of states and $A$ represents the set of actions that the agent can take.
The state transition function $T(s, a, s')$ describes the probability of transitioning from state $s$ to state $s'$ when taking action $a$.
The reward function $R(s, a, s')$ defines the immediate real-valued reward $r \in \mathcal{R}$ an agent receives when transitioning from state $s$ to state $s'$ by performing action $a$.
Finally, the discount factor $\gamma$ discounts future rewards when calculating the value of actions.

In order to solve an MDP, an RL agent interacts with the environment to derive an optimal policy $\pi^*$, which is a mapping from states to action probabilities such that behavior according to it maximizes the expected cumulative reward over time.
The \textit{value function} of a state $s$ under a policy $\pi$, denoted as $v_\pi(s)$, is the expected return when starting in state $s$ and following policy $\pi$ thereafter.
Here, $\mathbb{E}\pi$ denotes the expected value of a random variable given that the agent follows policy $\pi$. The function $v\pi(s)$ is the state-value function for policy $\pi$.

\begin{equation}
  v_\pi(s) = \mathbb{E}_\pi[G_t | S_t = s], \text{ for all } s \in S,
\end{equation}
where $G_t = \sum_{k=0}^\infty\gamma^kR_k$ represents the expected discounted return. The value of performing an action $a \in A$ while in a state $s \in S$ is defined as:

\begin{equation}
q_\pi(s,a) = \mathbb{E}_\pi[G_t | S_t = s, A_t = a], \text{ for all } s \in S \text{ and } a \in A.
\end{equation}

An optimal policy $\pi^*$ provides state-action associations that maximize the expected return or utility:

\begin{equation}
\label{eq:expected_return}
q_{\pi^*}(s,a) = \sum_{s',r}p(s',r|s,a)[r + \gamma \max_{a'}q_{\pi^*}(s',a')].
\end{equation}

The agent learns the optimal policy by interacting with the environment, receiving feedback in the form of rewards, and updating its value estimates for state-action pairs. In temporal difference (TD) learning, the value estimates are based on the difference between the expected and the observed value:

\begin{equation}
\label{eq:value_update}
v(s) \leftarrow v(s) + \alpha[R_{s'} + \gamma v(s') - v(s)],
\end{equation}
where $\alpha$ represents a learning-rate parameter. This operation updates the value associated with the state $s \in S$ as soon as the new state $s' \in S$ is reached, by computing the difference between predicted and observed values. Combining equations \ref{eq:expected_return} and \ref{eq:value_update} results in a form of TD learning called Q-learning, which can be expressed as

\begin{equation}
\label{eq:q_learning}
q(s,a) \leftarrow q(s,a) + \alpha[R(s,a) + \gamma \max_{a'}q(s',a') - q(s,a)].
\end{equation}

Building upon the TD update of value functions, we propose to derive appraisal computations that permit an assessment of events in connection with an agent's objectives and cognition.
Our model formalizes four appraisals out of the 14 from Scherer's CPM within the RL formalism: suddenness, goal relevance, conduciveness, and power.
We selected these four based on a minimal set that can be used to differentiate between emotions that are reportedly important and prevalent in interaction, such as joy, irritation, or boredom.
Eventually, other checks should be implemented for an accurate and extensive model.

\emph{Suddenness} is a part of the relevance appraisal in the CPM, and plays a role in checking how novel an event is. Building on the appraisal criterion of novelty from the CPM, we define suddenness as the frequency at which a transition into a state $s'$ occurs, given a previous state $s$ and action $a$ taken in it by the agent.
On a computational level, this will be represented as the relative frequency of the state's visitation, determining the level of suddenness of the event.
States to which the simulation transitions more frequently, given a previous state, are considered less sudden, and conversely, infrequent visitation results in greater suddenness appraisal of the event. To compute this, we introduce a suddenness measure $A_{s}$ defined as:



$$ A_{s} \propto 1- \frac{\hat{T}(s,a,s')}{\sum_{s'' \in S}\hat{T}(s,a,s'')}. $$
Here, $\hat{T}$ is a \emph{world model}, an approximation of $T$, based on the agent's accumulated experiences in its environment.

\emph{Goal relevance} is also checked during relevance appraisal in the CPM, and checks how relevant an event is, given the agent's current goal. While some goals are fairly general (e.g., survival), in interactive tasks, the goal relevance of an event can be related to the user's goals.
Generally, highly goal-relevant events elicit stronger emotional reactions than those less relevant to the agent's objectives. In our computational framework,
we operationalize the goal relevance to be proportional to the magnitude of the TD error observed during value prediction updates.
The reasoning for this choice is that the TD error focuses the agent's attention on events pertinent to the agent's goal via the learned utility function $q$, signaling that something in the environment has happened that impacts how the goal can be reached.
Both negative and positive implications of an event may be considered goal relevant:

$$ A_{gr} \propto \lvert \alpha[R(s,a) + \gamma \maxa_{a'}q(s',a') - q(s,a)]\rvert.$$
The intuition of this equation is that goal relevance is not an inherent property of the event, but a result of the agent's cognition computing the importance of the event to the eventual outcome that has relevance for the agent.

\emph{Conduciveness} appraisal is part of the implication assessment in the CPM, and checks if the event facilitates the attainment of the agent's goal.
It's essential to note that the intrinsic nature of events doesn't label them as conducive or obstructive. Instead, an agent's past experiences that associate these events with positive or negative outcomes play a role. 
In our computational model, conduciveness is represented not merely by the direction of the discrepancy between anticipated and actual outcomes, but also by the scale of that difference. We've quantified this by standardizing its values between 0 and 1. Therefore, a positive TD error indicates that the actual outcome was better than expected, and how much it surpassed expectations. Correspondingly, 1 is a very conducive event, indicating a considerable positive disparity. Conversely, a negative TD error might induce differing levels of negative emotions based on its value, with 0 representing a highly unconducive event, signifying a significant negative discrepancy. A value of 0.5, meanwhile, represents a neutral event, reflecting an event outcome that met the initial expectations. 
Again, an event does not have an intrinsic conduciveness, but it depends on the value update carried by the cognition of the agent, given existing expectations about the environment and the goals of the agent.

Goal conduciveness is therefore defined as:

$$ A_{gc} = min(max(\Delta, -1), 1) *0.5+0.5$$
where $\Delta$ is the TD error of the event (see Eq \ref{eq:q_learning}): $\Delta(s,a) = \alpha [R(s,a) + \gamma \max_{a'}q(s',a') - q(s,a)]$.

\textit{Power} appraisal is part of the coping assessment in the CPM.
It evaluates the agent's ability to impact the result of an event.
For instance, in our example in the introduction, the experienced user has power because they possess the knowledge to address the error, whereas the novice user lacks this capability.
In our model, power is based on the agent's capacity to choose between useful and non-useful actions in a given state.
If there are differences in the q values associated with different actions, the agent is presumed to have power to influence the event's outcome.
Conversely, if the q values associated with alternative actions are identical or if there is only one possible action to choose from, the agent is not considered to have power.
We quantify power as the difference between average q values and the minimum q value at a state, with higher values indicating a greater sense of power:

$$ A_{p} \propto \frac{\sum_a{q(s')}}{\lVert a \lVert} - \mini_{a'}q(s'). $$







Our model predicts four appraisal checks as a result of a state transition event.
This vector of four scalar values is then classified for predicting intensities of select modal emotions.
To this end, a Support Vector Machine (SVM) classifier was trained to predict modal emotions from appraisal patterns.
The mapping of patterns to modal emotions was adapted from an existing table \cite{scherer2001appraisal_book}, where, given a modal emotion, each appraisal check was given an intensity on a nominal scale.
For instance, the modal emotion ``joy'' is associated with high suddenness, high goal relevance, positive goal conduciveness, and medium power.
Because our model outputs scalar values for the appraisal checks, rather than words, we transformed the nominal scaling used in the table into distributions of values, as shown in Table \ref{tab:quantitave}.
The reason for using a distribution rather than exact numbers is that nominal values of appraisals, such as ``low'' or ``very high'' would be difficult to match to an exact number, leading to difficulties in the classifier.
As an example, values for low appraisal are distributed half-normally (denoted as $\mathcal{N}, x \geq 0$), with mean $\mu = 0$ and standard deviation $\sigma = 0.1$.
As a result of training the classifier on the theoretical data extracted from the original table, it can predict intensities of modal emotions from scalar appraisal profiles that are generated from our RL appraisal agent.
These profiles, adapted from \cite{scherer2001appraisal_book}, are shown in Table \ref{tab:appraisal}.

\begin{table}
\begin{center}
\caption{Mapping of nominal appraisal values to scales.}
\label{tab:quantitave}
\begin{tabular}{l|l}
  obstruct  & $\mathcal{N}, x \geq 0, \mu = 0, \sigma = 0.05$ \\
  very low  & $\mathcal{N}, x \geq 0, \mu = 0, \sigma = 0.05$ \\
  low       & $\mathcal{N}, x \geq 0, \mu = 0, \sigma = 0.1$  \\
  medium    & $\mathcal{N}, \mu = 0, \sigma = 0.05$  \\
  high      & $\mathcal{N}, x \leq 1, \mu = 1, \sigma = 0.1$  \\
  very high & $\mathcal{N}, x \leq 1, \mu = 1, \sigma = 0.05$ \\
  open      & $ \text{U}(0,1)$\\
\end{tabular}
\end{center}
\end{table}

\begin{table}
\begin{center}
\caption{Appraisal patterns for selected emotions and appraisals. Goal rel. = goal relevance, Conduc = conduciveness, obs. = obstrcut.}
\label{tab:appraisal}
\begin{tabular}{l| l l l l l l l l l l l}
  & Suddenness & Goal rel. & Conduc. & Power \\
  \hline
Happiness        & low        & medium         & high          & open \\
Joy              & high/med   & high           & very high     & open \\
Pride            & open       & high           & high          & open \\
Boredom          & very low   & low            & open          & medium \\
Fear             & high       & high           & obs.      & very low \\
Sadness          & low        & high           & obs.      & very low \\
Shame            & open       & high           & obs.      & open \\
Anxiety          & low        & medium         & obs.      & low \\
Despair          & high       & high           & obs.      & very low \\
Irritation       & low        & medium         & obs.      & medium \\
Rage             & high       & high           & obs.      & high \\
\end{tabular}
\end{center}
\end{table}

This study raises a potential discrepancy with Scherer's appraisal theory regarding the suddenness appraisal of shame. While Scherer's table categorizes shame as low in suddenness, our analysis suggests that this emotion should be classified as open. This discrepancy arises from the fact that shame can result from sudden events, such as making a mistake in front of others, but it can also be a more persistent emotion related to a person's self-identity and self-worth. As a result, we argue that the suddenness appraisal of shame should be more flexible and context-dependent. Other than that, the appraisal values in Table \ref{tab:appraisal} are the same as Scherer's table. It is an extraction of the original table, only including the appraisals and emotions that are modeled in our two experiments.


\section{Experiments}
\subsection{General Method}

Our model was validated via two vignette studies, requiring human participants to read short narratives and evaluate the emotions experienced by the protagonists.
These stories portrayed the protagonists interacting with various scenarios in which they were likely to experience a specific emotion.
Each story was constructed using the principles of appraisal theory to ensure the elicitation of the desired emotional response.
For example, to induce an emotion associated with a low power appraisal, we developed a story highlighting the protagonist's lack of control, consistent with the appraisal profile of that emotion (see Table \ref{tab:appraisal}).
After reading each vignette, participants were asked to provide intensity ratings for the emotions they believed the protagonist would experience.
This approach facilitated the comparison of the human intuitive understanding of emotion and the model's predictions.

\subsubsection{Materials}

We developed 11 narratives, each designed to elicit one of the 11 different emotions outlined in Table \ref{tab:appraisal}.
Each vignette, ranging between 90 and 200 words, depicted a protagonist interacting with technology in a way that elicits specific emotional responses.  
We crafted the content of each story to align with the appraisal profile corresponding to the targeted emotion.
For instance, in a narrative aimed at eliciting fear, the event portrayed was sudden and highly relevant to the protagonist's goals but presented significant obstacles and offered the protagonist little power to alter the outcome.

\subsubsection{Procedure}

Participants were recruited online and directed to a website hosting the vignettes.
After reading each story, they completed a questionnaire asking them to rate the intensity of various emotions they believed the protagonist would experience on a scale from 0 (not at all) to 10 (extremely).
To mitigate the potential influence of story sequence on the results, we employed a Latin square counterbalancing design, ensuring the presentation order varied across participants.

\subsubsection{Data Analysis}

Data collected from the online experiment were aggregated to calculate the mean rating for each emotion in every story.
Before this, each participant's responses to each story was standardized to minimize the effect of different individuals using the scales differently.
We utilized multilevel modeling (via the \texttt{lme4} package for R) to test the hypothesis that the story influenced emotion ratings.
In addition, in the figures showing human and model data, we include approximate 95\% confidence intervals, which can be used to assess the spread of human responses but are not formal hypothesis tests.
To facilitate comparison between human data and model predictions, we rescaled the human ratings to range from 0 to 1.

For generating model predictions, we designed 11 MDPs to represent the key events in each story.
All stories and the associated MDPs are detailed in Appendices~\ref{app:story} and~\ref{app:MDP}. 
A tabular RL agent was trained using Q-learning to converge on a policy for each MDP.
From the converged models, we computed four appraisal measures using the equations outlined in the previous section.
The resulting appraisal vectors were classified into modal emotion probabilities using a support vector machine (SVM).
 

The training and testing data for the classifier were simulated data derived from Scherer's table, see Table \ref{tab:quantitave} and \ref{tab:appraisal}. 
The classifier was trained using a Support Vector Machine (SVM), a supervised machine learning algorithm used for classification. One of the essential hyperparameters for SVM is the penalty parameter $c$ which controls the trade-off between maximizing the margin and minimizing classification error.
In determining the optimal $c$ value, we looked to human performance data as a reference. Specifically, we derived a mapping from observed human performance metrics to potential $c$ values. To illustrate, 
when considering a specific human precision 
- which reflects the probability that participants correctly identified a targeted emotion, we determine the corresponding $c$ value. Using this value within the SVM on simulated data would produce a similar precision in its prediction confidence of the targeted emotion. Through training multiple SVM classifiers over a spectrum of $c$ values on the simulated data, we selected the $c$ value that best mirrored the human precision, denoting as $c_{\text{mean}}$. 

Recognizing the importance of individual variability in human performance, we didn't solely rely on one SVM classifier to emulate all participants. We computed the variance in human precision, which in turn informed the variance of the $c$ value, labeled as $c_{\text{var}}$. Consequently, for each of the n participants in our experiments, we created an SVM classifier. The 
$c$ value for each classifier was drawn from a normal distribution defined by $c_{\text{mean}}$ and $c_{\text{var}}$. Importantly, to avoid overfitting, the value $c$ was not fitted to maximize our emotion predictions with the human data, but merely to align the classifier's precision with human precision.



\subsection{Experiment 1}



In the first experiment, we selected seven stories, each targeting a distinct modal emotion.
These emotions were happiness, joy, pride, boredom, sadness, shame, and fear.
The selection comprises three positive emotions (happiness, joy, and pride), three negative emotions (sadness, shame, and fear), and one neutral emotion (boredom).
Each emotion could be predicted by appraisal theory and therefore by our model based on the four appraisals we implemented computationally (see Table \ref{tab:appraisals_exp1}.



\begin{figure}[!t]
    \centering
    \includegraphics[width=7cm]{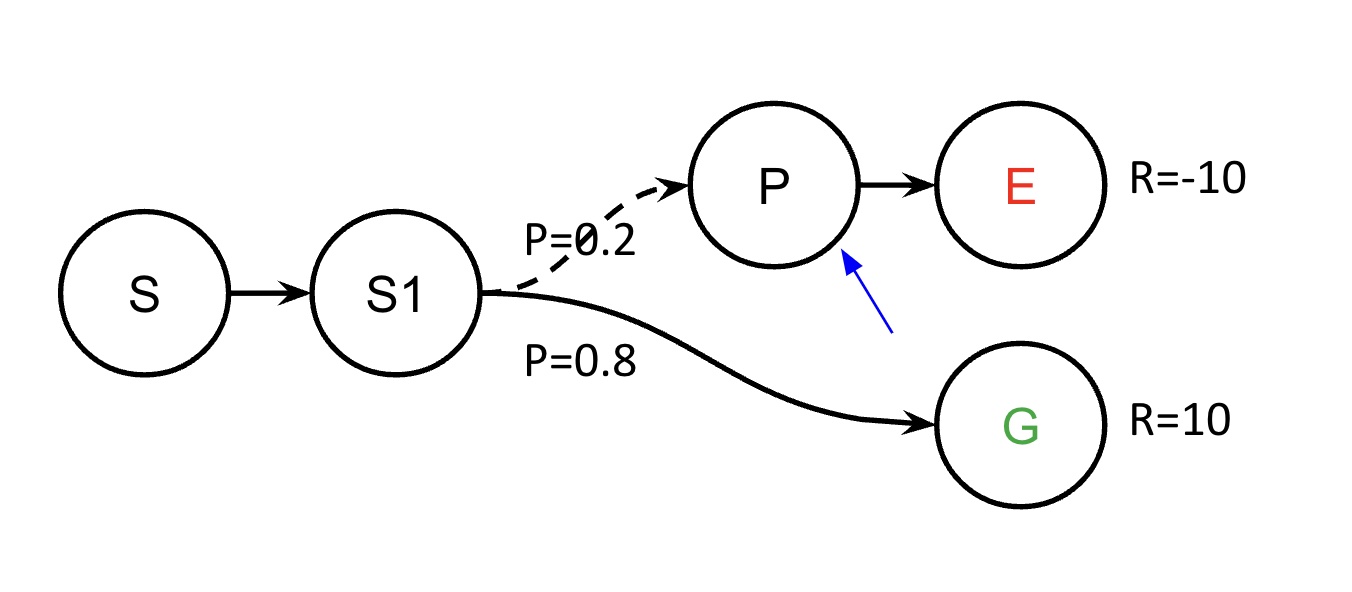}
    \caption{Fear MDP}
    \label{fig:fear_mdp}
\end{figure}

Figure \ref{fig:fear_mdp} illustrated an example MDP used in the experiment.
It shows how we model the story for fear, where the individual with no computer skills is taking an online exam when the internet suddenly goes off.
In the MDP, we describe the goal of the participant as a state G, which produces a positive reward.
Taking the only available action in state S1 often leads to the goal, leading the RL agent to have an expectancy of a stable internet connection.
However, sometimes the internet connection fails, resulting in the MDP transitioning to a problem state P, where the only available action is to move forward to a negatively rewarding error state E.
The appraisal analysis occurs at the onset of the problem, i.e., when the agent transitions from S1 to P.
This unexpected event carries a negative TD update due to the expected error state that follows the problem state.
The agent has no power to cope with the situation, because there are no alternative actions.
The resulting appraisal profile corresponds to fear.
Table \ref{tab:appraisals_exp1} shows all model-generated appraisal profiles.


\begin{table}
\begin{center}
\caption{Appraisal profiles generated by our model.}
\label{tab:appraisals_exp1}
\begin{tabular}{l|cccc}
  & Suddenness & Goal rel. & Conduc. & Power  \\
  \hline
  Happiness     & 0   & 0.67 & 0.83 & 0.95  \\
  Joy           & 0.8 & 1    & 1    & 0     \\
  Pride         & 0.5 & 1    & 1    & 0.1   \\
  Boredom       & 0   & 0    & 0.5  & 0.6   \\
  Fear          & 0.8 & 1    & 0    & 0     \\
  Sad           & 0.2 & 1    & 0    & 0     \\
  Shame         & 0.79& 1    & 0    & 0.5   \\
\end{tabular} 
\end{center}
\end{table}

\begin{figure}[!t]
    \centering
    \includegraphics[width=8.5cm]{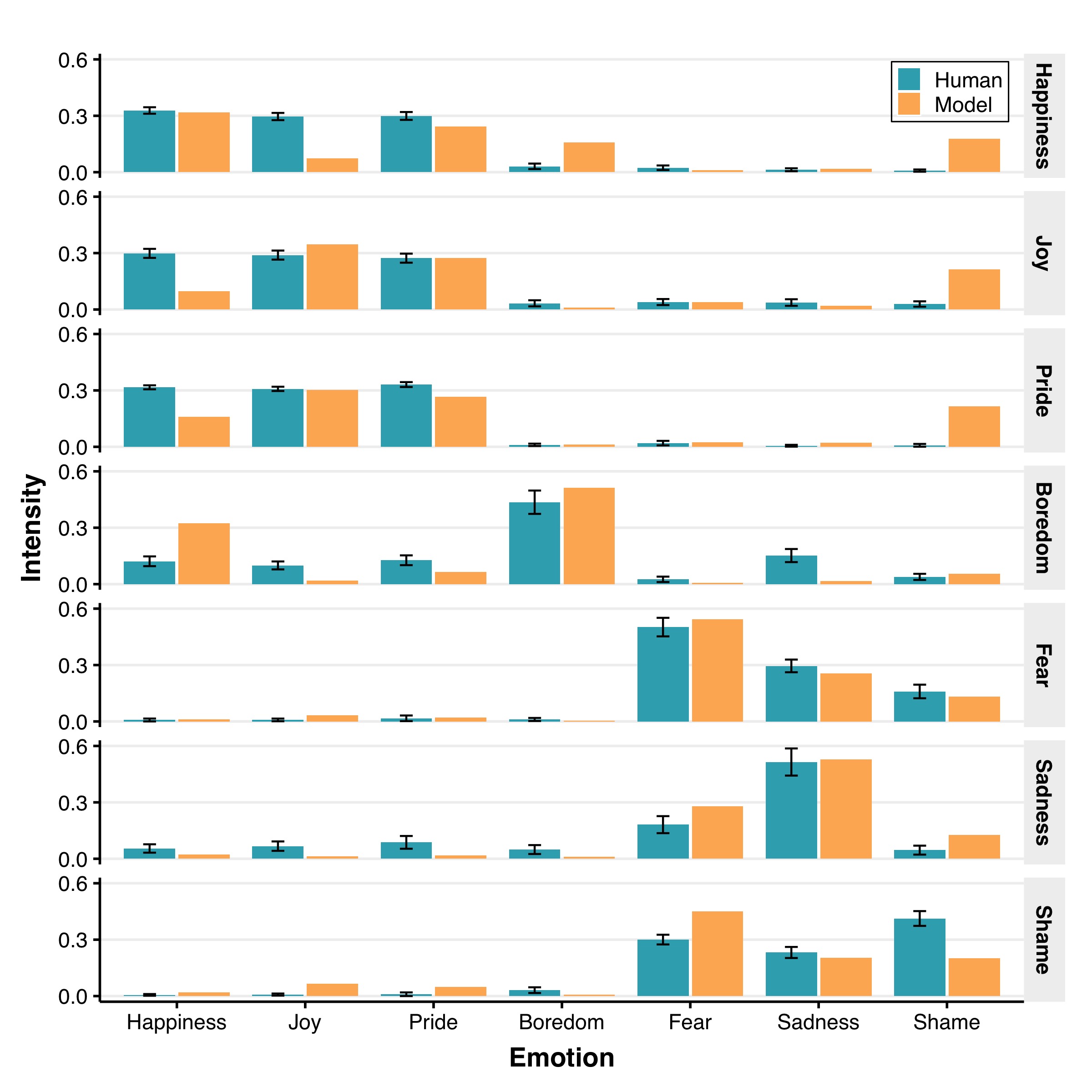}
    \caption{Comparison of human and model predicted emotional ratings for each vignette. The bars represent the mean ratings of each emotion by the participants (in blue) and the model's predictions (in orange) for each of the seven stories. The clear division between positive and negative emotions across stories is evident in both human ratings and model predictions. The error bars indicate approximate 95\% confidence intervals.}
    \label{fig:exp1_human}
\end{figure}

The first experiment was conducted as an online study with 42 participants (37 women, 5 men) and a mean age of $39.5$ ($sd = 12.8$).
Their average ratings for the seven vignettes, as compared to the model's predictions, are presented in Figure \ref{fig:exp1_human}.
There was a statistically significant interaction effect on the rating between emotion and story, meaning that the stories impacted emotion ratings, as hypothesized, $F(36, 2009) = 176, p < .001$.
A notable observation from these data is the clear distinction between positive and negative emotions.
Stories designed to evoke happiness, joy, or pride resulted in high ratings for all three emotions.
Similarly, in vignettes intended to elicit fear, sadness, or shame, these emotions were rated more prevalently.
For the vignette aiming to induce boredom, participants correctly identified this as the most probable emotional response of the protagonist.
Overall, our model achieved a reasonable degree of fit to the data ($c_{\text{mean}} = 0.0032, c_{\text{var}} = 0.0002, R^2 = 0.65, \text{RMSE}=0.09$).


The results indicate a human tendency to identify multiple co-occurring emotions within the same scenario, a finding that has been made frequently with studies that permit free rating of multiple modal emotions \cite{jokinen2015emotional,jokinen2019elicitation,saariluoma2014emotional,saariluoma2015appraisal}.
For instance, the story that was designed to evoke happiness also resulted in high ratings for joy and pride, suggesting that these emotions, while distinct, are often experienced together.
Similarly, negative emotions such as fear, sadness, and shame were all rated highly in their respective scenarios.
By fitting our model's sensitivity parameter $c$, we replicated this phenomenon, effectively accounting for the co-existence of multiple emotions within a single narrative.

However, while our model demonstrated a satisfactory degree of fit in terms of error, the somewhat lower $R^2$ results from not much variance to explain due to the relatively uniform ratings for all positive or negative emotions within each story.
In light of this, we designed a follow-up experiment to constrain the participants' freedom in assigning intensity ratings to each emotion.
This permits uncovering if a single modal emotion is more probable in each narrative context.
Similarly, the sensitivity of the model can be limited to align its predictions with the more constrained human ratings.

\subsection{Experiment 2}

\begin{figure}[!t]
    \centering
    \includegraphics[width=8.5cm]{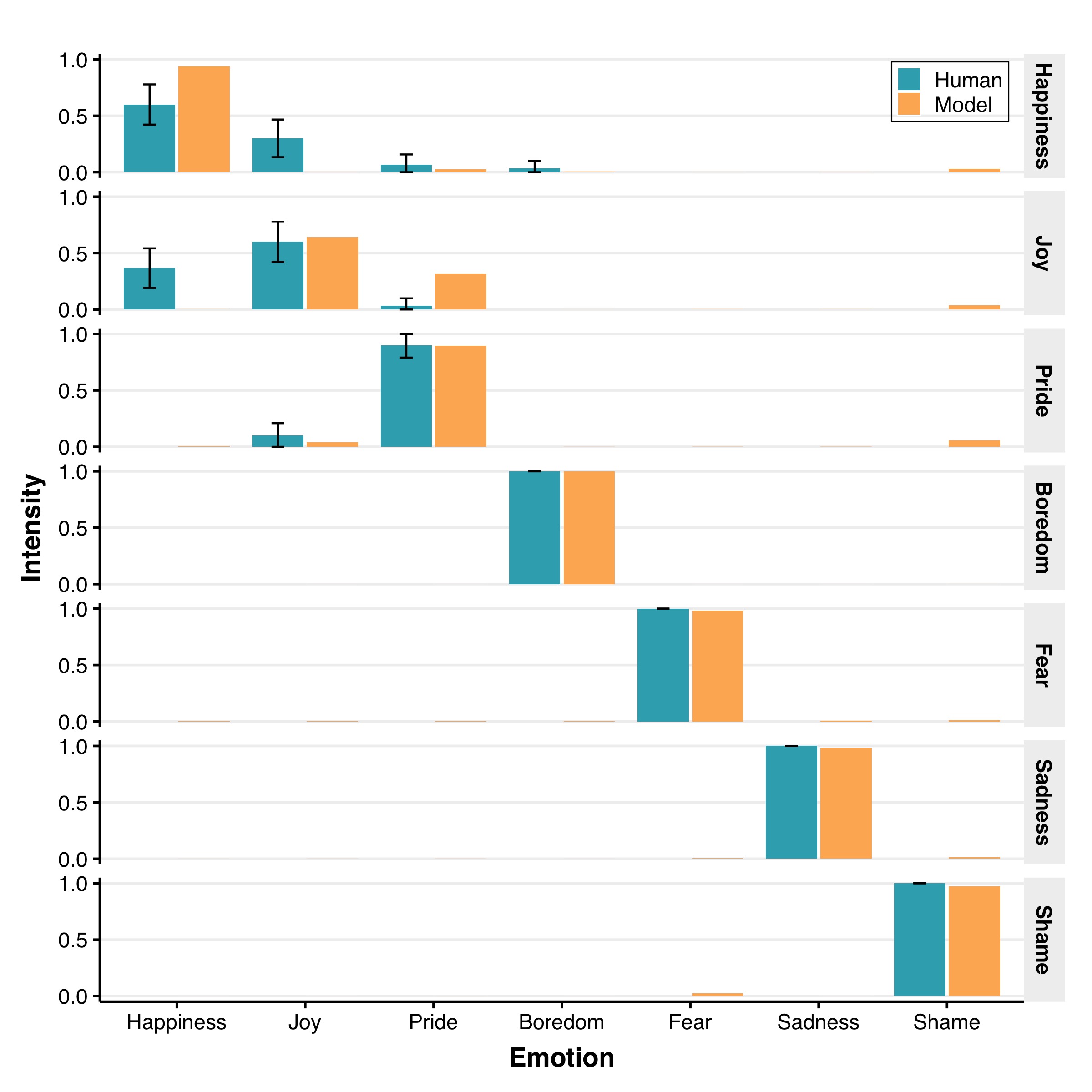}
    \caption{Comparison of the probabilities of each emotion being selected as the most prominent in a given story by human participants (blue bars) against the model's predicted intensity (orange bars). Each cluster of bars represents a story, with the seven emotions on the x-axis and the probability or intensity on the y-axis. The error bars indicate approximate 95\% confidence intervals.}
    \label{fig:exp2_human}
\end{figure}

The second experiment retained the materials from the first experiment, but modified the procedure to involve selecting a singular, most prominent emotion for each story.
The selection was done only after reading each story one by one.
In the final stage of the trial, the participants could again see all stories, and had to assign one modal emotion to each story, using each emotion in the process.
New participants ($N = 30$) were recruited online, with a mean age of $35$ ($sd = 5$), 26 women and 4 men.
The probabilities of each emotion being selected as the most prominent in a given story were compared to the model's predicted intensity, as visualized in Figure \ref{fig:exp2_human}.

There was a statistically significant interaction effect on the rating between emotion and story, meaning that the stories impacted emotion ratings, as hypothesized, $F(36, 1421) = 153, p < .001$.
The participants can be seen to associate the intended modal emotion with the corresponding story successfully.
Model fit was good, $c_{\text{mean}}=0.014,c_{\text{var}}=0.0056, R^2=0.92, \text{RMSE} = 0.09$.

The outcome of the second experiment demonstrates our model's ability to predict the most likely modal emotion within stereotypical contexts involving technology interactions.
The parameter $c$ can be used to calibrate our model's sensitivity to a particular modal emotion vs. a wider array of emotions and their intensities.
There was a marked consensus between human participants and the model regarding the predominant modal emotion in each story, thanks to carefully designed narratives and their corresponding MDPs, which implemented the four appraisals: suddenness, goal relevance, conduciveness, and power.
Despite the added complexity of the task, requiring a mapping of each of the seven emotions to a unique story, the participants and the model performed well.
This outcome reinforces the notion of an inherent human capacity for ``intuitive appraisal'', enabling us to model and predict others' emotions effectively in everyday situations \cite{saxe2017formalizing}.

These findings open up a question: although humans and the model accurately differentiate between starkly contrasting emotions, such as happiness and sadness -- a capability that aligns with the predictions of the appraisal theory and the Component Process Model (CPM) -- can they distinguish between nuanced emotional states that diverge on just one appraisal?
This question prompts our third experiment, where we aim to explore this capability to discern between closely related emotions that, nonetheless, differ on a key appraisal.
If humans and our model can indeed model emotions according to the CPM, even a minor change in the power appraisal of the protagonist or the simulated user should yield a different modal emotion.

\subsection{Experiment 3}

In the third experiment, we developed new materials: four narratives were specifically crafted to elicit negative emotions.
These emotions -- anxiety, desperation, irritation, and rage -- could be evoked by adjusting one or two selected appraisals: suddenness and power.
These four emotions share certain traits in their appraisal profiles: all are elicited by events that are goal-relevant and obstructive (non-conducive).
However, they differ in their requirement for suddenness: desperation and rage necessitate a sudden event, whereas anxiety and irritation do not (see Table \ref{tab:appraisals_exp3}).
Furthermore, the appraisal of power enables differentiation between anxiety and desperation (where there is a lack of power) from irritation and rage (where there is power).
This design permits the implementation of four distinct emotional scenarios through the manipulation of just two appraisals.
If both human participants and our model can discern these subtleties, it would improve the plausibility of the CPM and our computational implementation of it.
All model-generated appraisal profiles of these four emotions are shown in Table \ref{tab:appraisals_exp3}.

\begin{table}
\begin{center}
\caption{Appraisal profiles generated by our model.}
\label{tab:appraisals_exp3}
\begin{tabular}{l|cccc}
                & Suddenness & Goal rel. & Conduc. & Power  \\
                \hline
  Anxiety       & 0.2 & 1 & 0 & 0  \\
  Despair       & 0.81 & 1 & 0 & 0  \\
  Irritation    & 0.2 & 1 & 0 & 0.53  \\
  Rage          & 0.8 & 1 & 0 & 0.6  \\

\end{tabular}

\end{center}
\end{table}

Incorporating elements from the first two experiments, the procedure for the third experiment asked human participants to both freely rate the emotions after reading each story (as in Experiment 1) and later, associate each story with the most probable modal emotion (as in Experiment 2).
The study comprised 29 online participants, with 26 women, 8 men, and a mean age of $M_{age} = 35.2$, $sd = 11.9$.

\begin{figure}[!t]
    \centering
    \includegraphics[width=8.5cm]{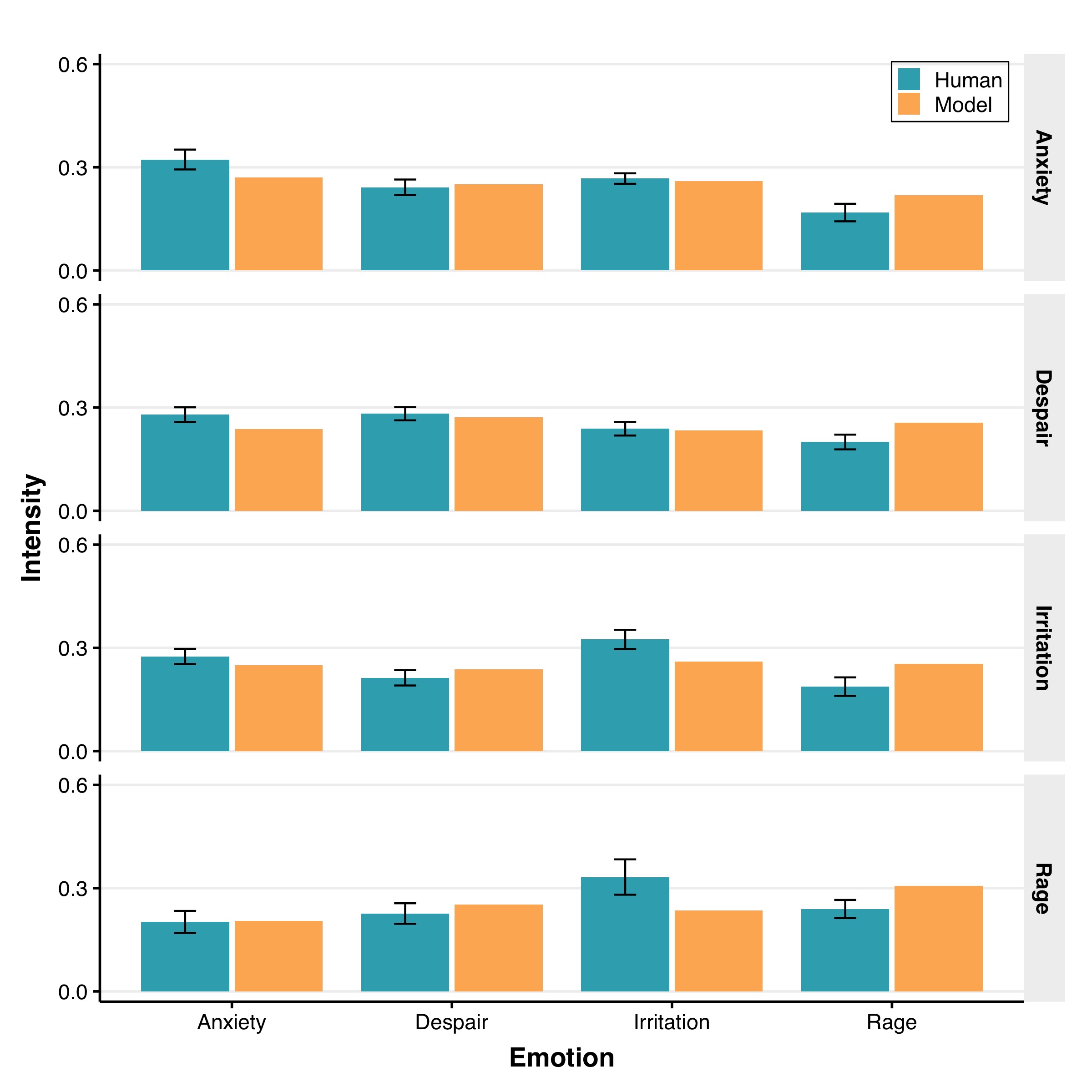}
    \caption{Comparison of human and model predicted emotional ratings for each vignette. The bars represent the mean ratings of each emotion by the participants (in blue) and the model's predictions (in orange) for each of the four stories. The error bars indicate approximate 95\% confidence intervals.}
    \label{fig:exp3_free}
\end{figure}

Figure \ref{fig:exp3_free} displays the average ratings for the four vignettes alongside the model's predictions.
There was a statistically significant interaction effect on rating between emotion and story, meaning that the stories impacted emotion ratings, as hypothetized, $F(9, 528) = 12, p < .001$.
Given that all stories suggested a similar negative emotion, the ratings exhibited limited variance.
The results were consistent with those of the first experiment, and our model could replicate the overall outcome, $c_{\text{mean}} = 0.0013, c_{\text{var}} = 0.0001, R^2 = 0.29, \text{RMSE} = 0.04$.
Despite the shared variance between model predictions and human data being relatively small due to the limited variance between the ratings, the model error was minimal, indicating a good fit between the model and human responses.

\begin{figure}[!t]
    \centering
    \includegraphics[width=8.5cm]{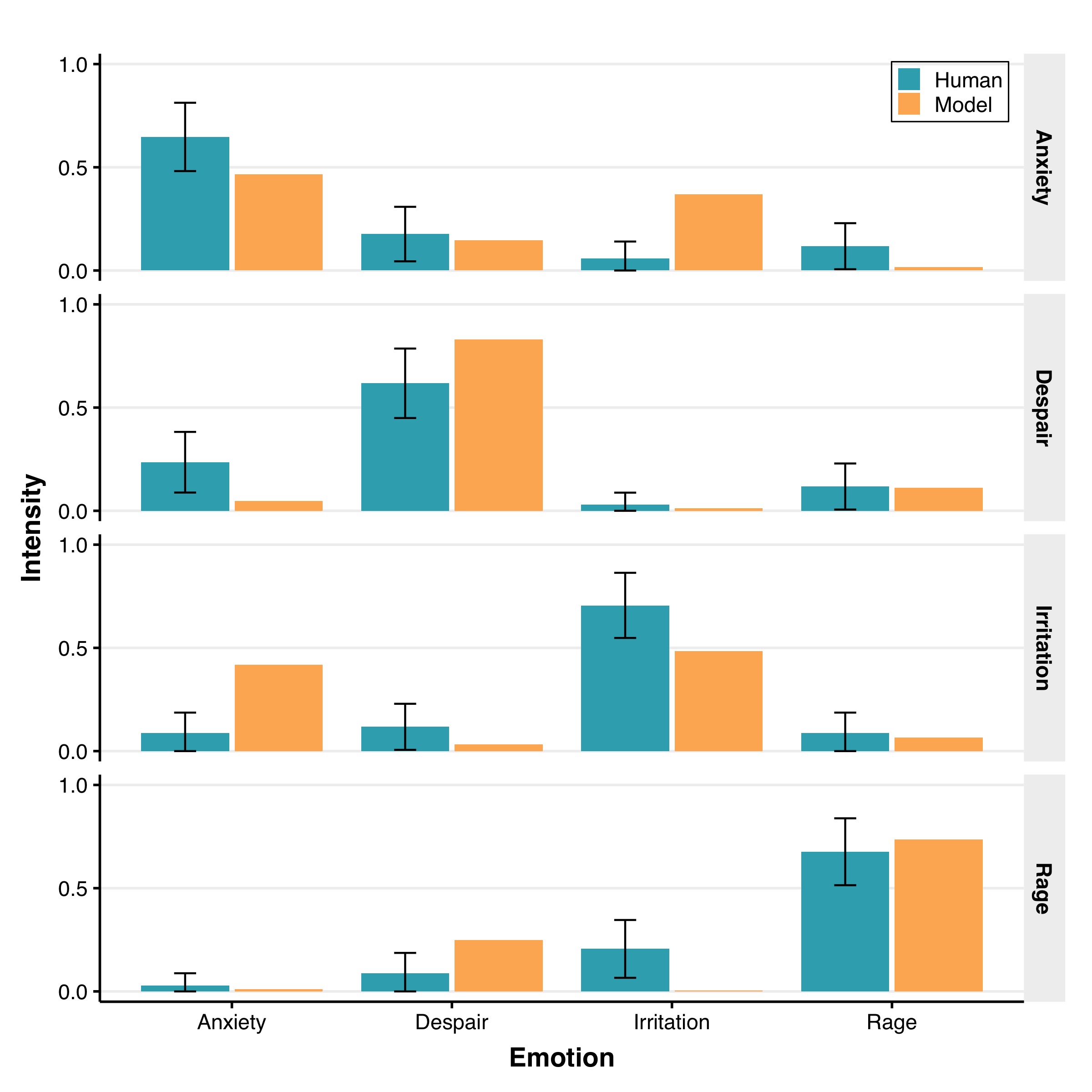}
    \caption{Comparison of the probabilities of each emotion being selected as the most prominent in a given story by human participants (blue bars) against the model's predicted intensity (orange bars). Each cluster of bars represents a story, with the four emotions on the x-axis and the probability or intensity on the y-axis. The error bars indicate approximate 95\% confidence intervals.}
    \label{fig:exp3_limit}
\end{figure}

In the forced-choice responses, where participants had to assign one unique emotion to each story, they were able to accurately identify the intended modal emotion, as shown in Figure \ref{fig:exp3_limit}.
There was a statistically significant interaction effect on rating between emotion and story, meaning that the stories impacted emotion ratings, as hypothetized, $F(9, 528) = 27, p < .001$.
The model demonstrated a reasonable fit with human data, $c_{\text{mean}} = 0.0034, c_{\text{var}} = 0.001, R^2 = 0.62, \text{RMSE} = 0.16$.
The shared variance increased as both humans and the model were compelled to be more sensitive to specific emotions.
While the model error was higher, the model was still able to predict the most salient emotion in a manner comparable to humans.

The findings from our third experiment provide more nuanced insight about both our model and human emotional perception.
In contrast to the clear emotional distinctions of the previous experiments, the third experiment tested the ability to discern between closely related emotions that differ in only one or two key appraisals.
Despite the increased complexity of this task, both human participants and our model were able to match emotions accurately with their intended modal emotions.

The outcome of the experiment supports the CPM approach to modeling human emotion, reinforcing the notion that emotions can be differentiated by varying the appraisals of power and suddenness.
The ability of our model to mimic human responses across all three experiments provides validity to our RL-based computational implementation of the CPM, even when dealing with subtler emotional differences.
It suggests that our model has effectively captured the underlying appraisal processes that govern emotional responses.

For an overview of the model fit across the three experiments, a summary is provided in Table \ref{tab:master}.
The results of the experiments also illuminate the flexibility of human emotional perception.
When given freedom, as in Experiment 1, participants displayed an ability to perceive multiple emotions in response to a single narrative.
This could be matched by our model by varying the sensitivity parameter $c$.
However, when constrained, as in Experiments 2 and 3, humans and our model could also pinpoint a single most salient emotion, even when the differences between emotions were subtly manipulated through appraisal variations.
This adaptability underscores the complexity of human emotion recognition and the efficacy of the CPM in modeling such processes.

\begin{table*}
\begin{center}
\caption{Master table.}
\label{tab:master}

    \begin{tabular}{|c|c|c|c|c|c|c|c|}
    \hline
      &Num. of &\multicolumn{2}{c|}{Human precision} & \multicolumn{2}{c|}{Parameter C} &
      Multiple &\multirow{2}{*}{RMSE}\\
      \cline{3-6}
      &participants &Mean &Variance &Mean &Variance &R-squared&\\
    \hline
      Exp1 & 42 & 0.40  &0.028  & 0.0032 &0.0002 & 0.65 &0.09\\
      \hline
      Exp2 &30  &0.87 &0.112 &0.014&0.0056&0.92 &0.09\\
      \hline
      Exp3 (free choice) &34  &0.29 &0.0067   &0.0013&0.0001& 0.29 & 0.04\\
      \hline
      Exp3 (forced choice)&34  &0.66   & 0.2238 & 0.0034 & 0.001 & 0.62 &0.16\\
      \hline
    \end{tabular}
    
\end{center}
\end{table*}





\section{General Discussion and Conclusion}
\subsection{General Discussion}

In this work, we propose a novel computational model that harmonizes elements of Reinforcement Learning and appraisal theory. This approach distinguishes our work from related work, facilitating the development of a comprehensive theoretical framework capable of generating behavioral trajectories based on set goals, individual capabilities, and the task environment. Consequently, it provides an enriched understanding of the cognitive substrates involved in emotional experiences during interactions.


The distinct advantage of our model resides in its capability to generalize across a wide range of appraisals and tasks, which can be represented as a Markov Decision Process (MDP). There exists a substantial body of work employing MDPs and RL for interaction modeling, encompassing areas such as visual search \cite{chen2017cognitive}, multitasking \cite{jokinen2021modelling}, and typing \cite{jokinen2021touchscreen}, among others. Significantly, our model can be applied to anticipate emotions in these interaction tasks.



Our model has certain limitations. Despite its capability to generalize across several appraisals, our model does not encompass all potential cognitive evaluations, thus omitting possible influences on emotional responses. For instance, while internal and external standards could be represented in an MDP model, we have excluded them here due to the added complexity of introducing multiple agents. Furthermore, representing appraisal urgency presents a challenge within the constraints of MDP models. In the future, we aspire to advance our model to predict emotions during various task stages, rather than solely at a specific state. This refinement represents an important trajectory for future research.


Looking forward, we aim to enhance our model by integrating a broader array of cognitive evaluations and exploring alternate learning paradigms. In parallel, we envisage transitioning our focus from controlled vignette studies to actual human interactions and emotional responses. 
Starting with vignettes remains a worthwhile approach in our study, as they provide a standardized, controlled platform that allows for consistently examining emotional responses across different scenarios. But they inherently lack the authenticity of real-life experiences.
This dual approach will not only address current limitations but also amplify the model's versatility and applicability, thereby providing deeper insights into the intricate interplay between cognition and emotion.

\subsection{Conclusion}
The introduction of our model marks a significant advancement in the field of emotion modeling, primarily due to its unique combination of appraisal theory and reinforcement learning (RL). This integration facilitates a more robust and nuanced understanding of emotional responses. By incorporating cognitive appraisal, our model acknowledges that emotions are not merely reactive, but rather are closely tied to our evaluations and interpretations of events. This positions our model to better simulate the subjective and highly personal nature of emotional experiences.

As we conclude, the future of emotion modeling, particularly within the realm of affective computing, is exciting. 
Our research represents one step forward, highlighting the interplay between cognitive processes, reward-based learning, and emotional experiences. 
As computational models grow more sophisticated and versatile, we anticipate a future where we can simulate human emotional responses with increasing accuracy, catering to various applications for human-computer interactions.

\bibliographystyle{IEEEtran}
\bibliography{references}

\newpage

\begin{IEEEbiography}[{\includegraphics[width=1in,height=1.25in,clip,keepaspectratio]{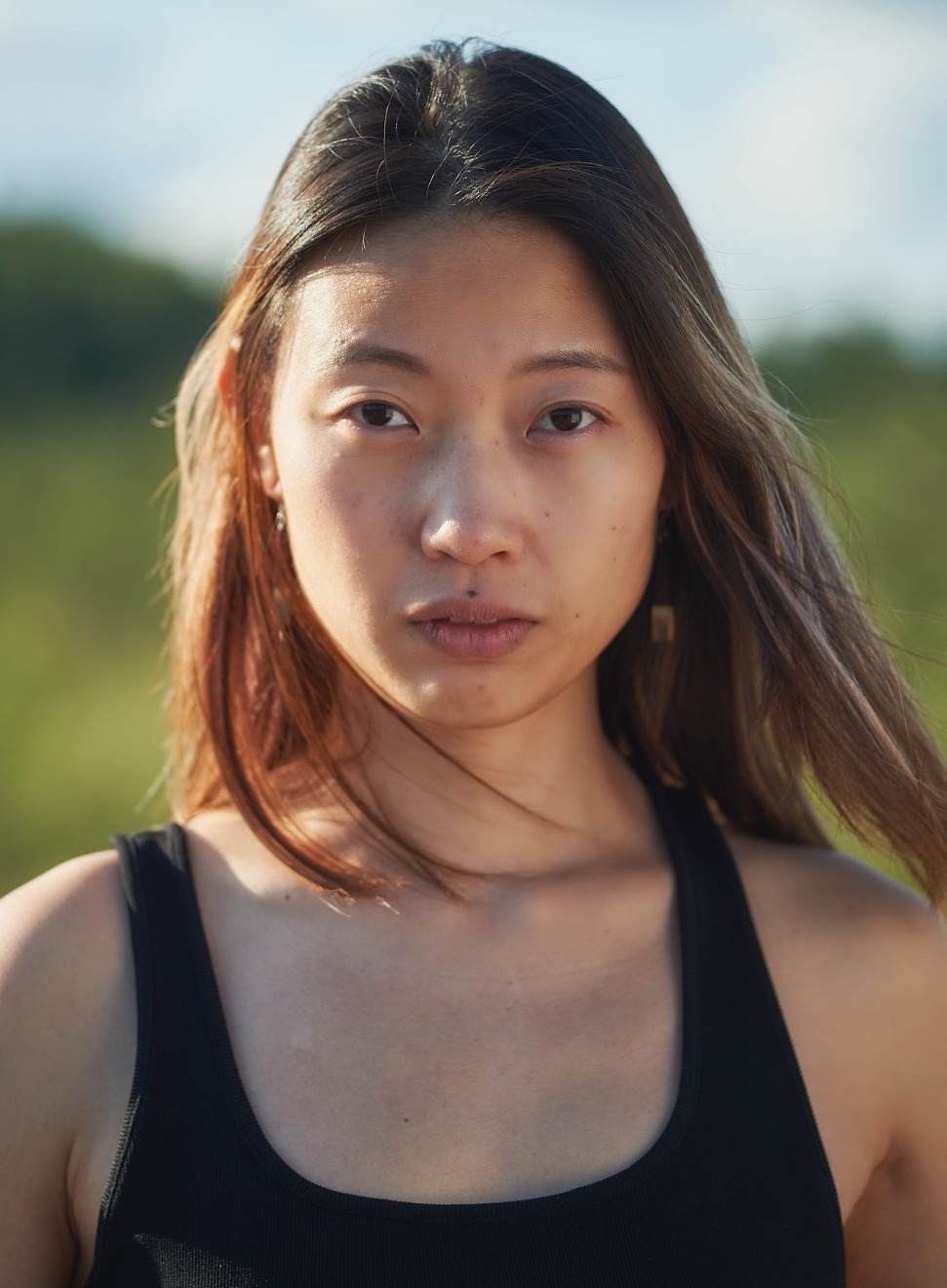}}]{Jiayi Zhang}
received the MSc degree in Human-Computer Interaction from Aalto University, Finland and Université Paris-Saclay, France in 2021. 
She is currently working towards the PhD degree with the Faculty of Information Technology in the university of Jyväskylä, Finland. Her research interests include emotion modelling and human-computer interaction.
\end{IEEEbiography}
\vspace{-12cm}

\begin{IEEEbiography}
[{\includegraphics[width=1in,height=1.25in,clip,keepaspectratio]{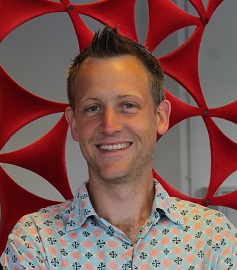}}]{Joost Broekens} is associate professor and head of the Affective Computing and Human-Robot Interaction lab at the Leiden Institute of Advanced Computer Science (LIACS), Leiden University. He is president emeritus of the Association for the Advancement of Affective Computing (AAAC). He is co-founder of Interactive Robotics and co-founder of Daisys. His research focuses on affective computing, in particular computational modelling of emotions in reinforcement learning and computational models of cognitive appraisal, and on human-robot interaction.
\end{IEEEbiography}
\vspace{-12cm}

\begin{IEEEbiography}
[{\includegraphics[width=1in,height=1.25in,clip,keepaspectratio]{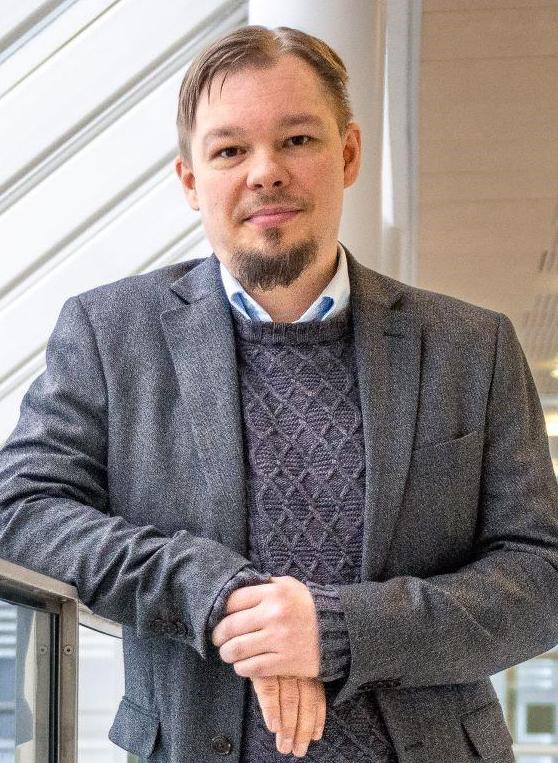}}]{Jussi P.P Jokinen} received his PhD in cognitive science from University of Jyväskylä, Finland, in 2015. He currently holds a position of Assistant Professor at University of Jyväskylä, Finland. His research covers a wide array of topics related to computational cognitive modeling and human-computer interaction.
\end{IEEEbiography}

\clearpage 
\appendix

\subsection{Stories}
\label{app:story}
Here are the stories that we have used for the vignette studies. 

For experiment 1 and 2:
\subsubsection{Happiness story}

John is a hard-working farmer and has made his farm prosper over the years. However, he wants some more free time in his life. For this reason, he has installed an automated system, which helps him monitor the soil condition and takes of the irrigation automatically.

The system worked well last year, greatly reducing his workload, but still resulting in comparable yields on the farm. Now John is more experienced with using the system. He can take his time to operate it and do the harvest. Now that everything is in place, John is assured of the harvest this year.

\subsubsection{Joy story}

Jack works as a developer in a software company. He has been developing software to help the university library with an e-book system. The project doesn't have a strict time constraint, so Jack can take his time for ensuring better product quality. Recently, an error in the developmental system has been causing problems with publishing it. Solving this is a crucial step: after it, the product is ready to be delivered. 

The problem is persistent, and Jack has tried out different solutions for days now, without really knowing what would work. One day, during the lunch break, he mentions his issue with a colleague. During the talk, John gets inspiration about a potential solution. Although he is in doubt, he gives it a try anyway. However, it works, and now the publishing procedure could be completed!

\subsubsection{Pride story}

James has worked as a developer in a software company for years. He wants a promotion and needs a project to prove himself. Now the opportunity has come. James is appointed to be in charge of a new project, which is to refine the university's e-book borrowing system for improving the students' experience. James takes the project seriously and wants to prove for himself to get the promotion. The university has high expectations for the project as well, because due to negative feedback about the system they desperately want to improve it. The university gives James sufficient time, wishing for high-quality work. 

James works very hard on the project. He collects feedback on the current system and tests several potential changes with students for their opinion. After spending weeks on it, James finishes a version that he is very satisfied with. And so is the university. After releasing James's version, the system's rating has increased significantly, and the e-book borrowing rates have gone up considerably. The librarian who is in charge of the e-service sent James a Thank-You email for his great work.

\subsubsection{Boredom story} 

David's been recruited as a website operator. His first task is to analyze several websites' daily visitors. He needs to transfer data to Excel and run a basic analysis on it. The task is simple and it is the only thing that he needs to do in his job. There are very rarely any problems, and David knows how to deal with any that might arise. David has been doing this work for weeks now. He has learned everything he can, and the work has become very repetitive and monotonous.

\subsubsection{Fear story}

Tom has a very important English language test coming. He needs a high score to apply to his dream university and this is his last chance to take the exam for this year's application. Because of Covid, the test will be taken remotely on their own computer, which requires those taking the test to have their cameras on all the time as an anti-cheating measure.

The exam day finally comes and Tom is well prepared. He successfully logs in to his account and starts the test. However, when he tries to load the next page, the website is suddenly stuck without any response. It looks like there might be a problem with Tom's internet connection. Tom is not good with computers or troubleshooting connection problems. He doesn't know what the problem is or how to fix it. Nevertheless, he needs the system to work again as soon as possible or he will have no time to answer the test. The test timer is still on but the page doesn't load.

\subsubsection{Sadness story}

Jim has been playing an online game for years. The game has introduced him to a lot of online friends, and it is important for him. He has basically grown up with the game and still plays it actively. However, with lots of competition in the gaming industry, the game is not as popular today as it has been before. Jim knows that the game will be shut down, as it has been non-profitable for some time now. When it shuts down, it will be completely unplayable. These days Jim has been saying goodbye to his game friends and quitting the game gradually.

\subsubsection{Shame story}

Peter is a highly-respected professor in Cybersecurity. One day he gets a suspicious email that his email account has been hacked. The message contains a link and encourages him to use it in order to rectify the situation. Peter immediately gets suspicious and realizes the risk of clicking the link. While the email does look professional, these types of messages are usually phishing and therefore dangerous. 

Peter thinks about reporting this email as phishing. But he is intrigued, so against better judgment he still clicks the link. Immediately, warnings are all over his screen and his email is locked for security reasons. This is his university working email that he needs to use all the time. Now he has to report himself to the IT department and admit to them that the "security professor got tricked'', in order to unlock his email.\\
\\
We used the emotions Anxiety, Despair, Irritation, and Rage in experiment 3. And they are the extensions of the Fear story in the previous experiments. These stories are very similar in content, with slight changes in the appraisals of suddenness and power. 

\subsubsection{Anxiety story}
John has a very important final exam coming, which will be done online using the school's exam system. John is a freshman and this is his first time using the system, and John is not so sure how it works.

As the exam starts, John successfully logs in to his school account and starts the test. But the loading is very slow when logging in, and the user interface feels very unfamiliar to him. He wants to explore some functions, but the system is extremely unresponsive. After answering all questions on the page, John clicks “Next Page" to proceed with the exam. However, the system appears to be stuck, and it just seems to be loading forever. 

John is not sure if this is normal, since the system was slow the whole time from the start. There should be more questions on the next page but the loading seems to be taking far too long. The timer is still on, and John is losing valuable exam time. He knows there must be something he can do, but this is his first time using the system.
\subsubsection{Despair story}
Frank has a very important final exam coming, which will be done online using the school's exam system. Frank is a freshman and this is his first time using the system, and he is not so sure about how to use it.

As the exam starts, Frank successfully logs in to his school account and starts the test. Although the exam interface looks unfamiliar, he manages to explore the functions a bit as the system is very responsive and has short loading times. The exam proceeds well and Frank has answered all questions on the page. Then Frank clicks “Next Page" to continue with the exam. However, the system unexpectedly responds that the exam is over. 

Frank is not sure what has happened, as he hasn't yet answered all the questions in the exam and he should have plenty of time left on the timer. Frank isn't sure if he has clicked the wrong button. He is fairly sure that the exam has now failed and there is probably nothing he can do about it. The whole semester's efforts on this course are gone if he has just submitted the exam himself.
\subsubsection{Irritation story}
Tom has a very important final exam coming, which will be done online using the school's exam system. Tom is a senior student and he has used the system multiple times before.

As the exam starts, Tom successfully logs in to his school account and starts the test. But the system is loading very slowly after login. Tom has had similar experiences before and he knows the system can be unresponsive like this. He just needs to be more patient. After answering all questions on the page, he clicks “Next Page" to proceed with the exam. However, the system appears to be stuck, and it just seems to be loading forever. 

Tom hasn't yet answered all the questions in the exam. The timer is still on, and Tom is losing valuable exam time. He knows that if the system crashed - as it looks like - he has to retake the exam at a later date. Tom can only wait to see if the system starts working and he can continue, or he has to prepare for the exam again later.
\subsubsection{Rage story}
David has a very important final exam coming, which will be done online using the school's exam system. David is a senior student and he has used the system multiple times before. David considers himself an expert in using it.

As the exam starts, David successfully logs in to his school account and starts the test. The system is very responsive and has short loading times, and David can focus on answering the questions. David is well prepared and he is sure that he can pass the exam this time. After answering all questions on the page, he clicks “Next Page" to proceed with the exam. However, the system unexpectedly responds that the exam is over. 

David had a similar experience before and he had already reported this issue to the school several times. He thought it would have been solved already, but here it is again. He knows that the system has crashed and he has to retake the exam at a later date. The rescheduled exam will cause David a lot of trouble out of blue, he has to change his vacation plans and has to prepare for the exam again later.

\subsection{MDP}
\label{app:MDP}
Here are the MDPs that we have used for each emotion. 

States (S): 
The states in the MDP are represented by circles. \( S \) denotes the starting state, while \( S_1 \) and \( S_2 \) are intermediary states. The goal state is represented by \( G \), and the error state is denoted by \( E \). Both \( G \) and \( E \) are terminal states, signifying the end of an episode.

Action (A): 
There are three possible actions: "forward", "action1", and "action2". When only one action is available to choose from, the agent defaults to choosing "forward" (denoted as \( frwd \)). When there are two available actions, the agent has the option to choose between "action1" (denoted as \( a_1 \)) and "action2" (denoted as \( a_2 \)).

Transition Probabilities (P): When not specified, the transition probability of taking an action to move from one state to the next is 1. If particular probabilities are outlined, executing an action in one state may lead to different states with the corresponding probabilities as depicted in the figures.

Rewards (R): Each transition into a state carries an associated reward. Transitioning into states \( S_1 \) or \( S_2 \) typically incurs a reward of generally -1, although exceptions may apply (e.g., in the emotion of happiness). The reward for reaching the goal state \( G \) is commonly 10, while the penalty for entering the error state \( E \) is -10. Again, the reward values in the termination states can differ in cases. 

Blue arrow: The blue arrow in the figures points to the state where the emotion occurs. This is the critical point at which we calculate the appraisals.

\subsubsection{Happiness}
Here, we wanted to emphasize that the farmer's life has been easier after having the automated system. Thus, in the training session, for the last 5 trials, the reward of transitioning into \( S_1 \) would drop from -3 to 0. And we wanted to emphasize that the farmer has good harvest results with the new system, also for the last 5 training trials, the reward is lifted from 7 to 10. In the testing episode, the agent is strategically designed to follow the sequence \( S \xrightarrow{frwd} S_1 \xrightarrow{a_1} G \). The appraisal analysis comes at the state \( S_1 \).

\renewcommand{\thefigure}{B.\arabic{figure}} 
\setcounter{figure}{0} 

\begin{figure}[h]
    \centering
    \includegraphics[width=8.5cm]{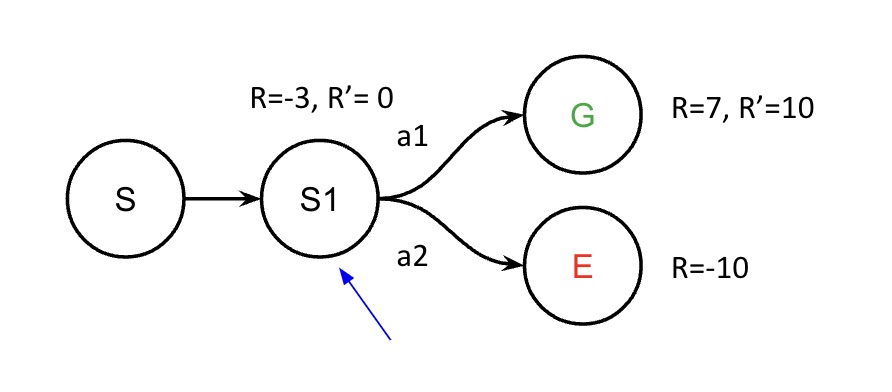}
    \label{fig:hap}
    \caption{MDP illustration for Happiness}
\end{figure}

\subsubsection{Joy}
Here is a story illustrating a person's journey to solve a problem, encountering numerous failures along the way, but experiencing joy upon success. At state \( S_1 \), when faced with the challenge, the agent is designed to fail 90\% of the time, leading to an error, while succeeding 10\% of the time, reaching the goal. In the testing episode, the agent is strategically designed to follow the sequence \( S \xrightarrow{frwd} S_1 \xrightarrow{frwd} G \). The appraisal analysis comes at the state \( G \).

\begin{figure}[H]
    \centering
    \includegraphics[width=8.5cm]{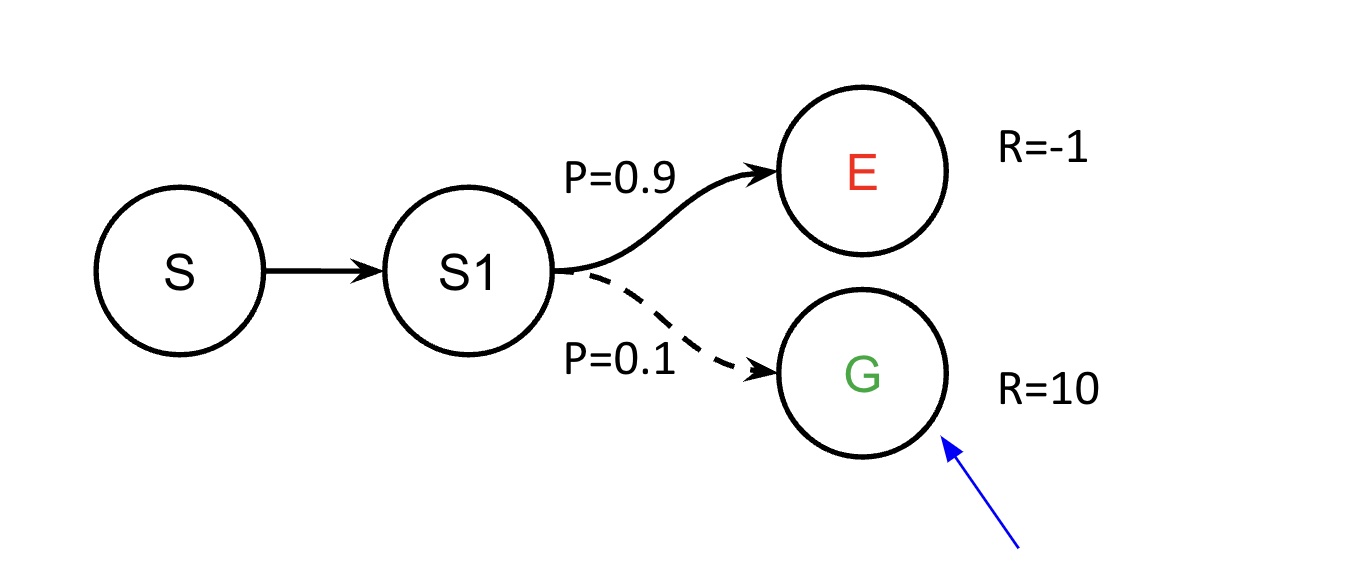}
    \label{fig:joy}
    \caption{MDP illustration for Joy}
\end{figure}

\subsubsection{Pride}
The pride story is about an individual who puts in extra effort to surpass their usual performance. The sequence \( S \xrightarrow{frwd} S_1 \xrightarrow{a_1} G \) represents the path that the agent typically takes, exerting less effort to achieve a normal result, with a reward of 5. If the agent chooses action \( a_2 \) at state \( S_1 \), it incurs a negative reward of -5, symbolizing additional work. However, this extra effort does not necessarily guarantee a superior outcome. Upon reaching state \( S_2 \), there is a 50\% chance that the agent will end with the normal result \( G \), and a 50\% chance that the agent will achieve a better result \( G+ \), with a reward of 10. In the final testing episode, the agent is designed to take \( S \xrightarrow{frwd} S_1 \xrightarrow{a_2} S_2 \xrightarrow{frwd} G+\). The appraisal analysis comes at the state \( G+ \).

\begin{figure}[H]
    \centering
    \includegraphics[width=8.5cm]{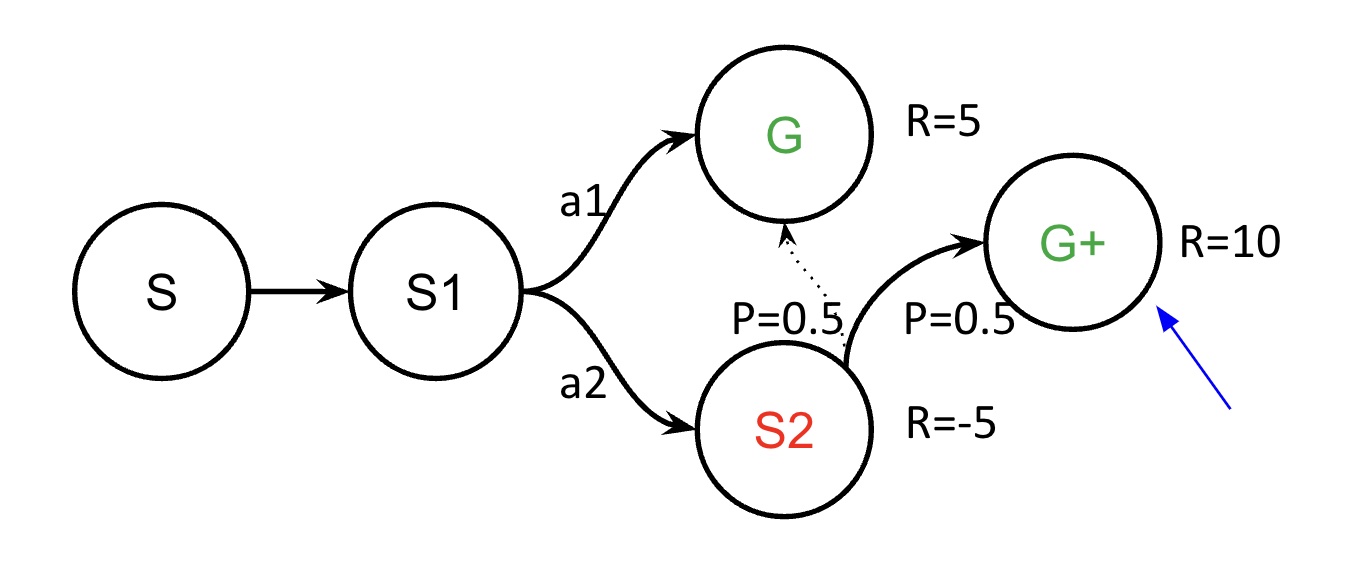}
    \label{fig:pride}
    \caption{MDP illustration for Pride}
\end{figure}

\subsubsection{Boredom}
The boredom story is about an individual engaged in repetitive tasks that yield limited rewards, in terms of both negative and positive rewards. In the final testing episode, the path taken by the agent is \( S \xrightarrow{\text{{frwd}}} S_1 \xrightarrow{a_2} G \). The appraisal analysis occurs at the state \( S_1 \).

\begin{figure}[H]
    \centering
    \includegraphics[width=8.5cm]{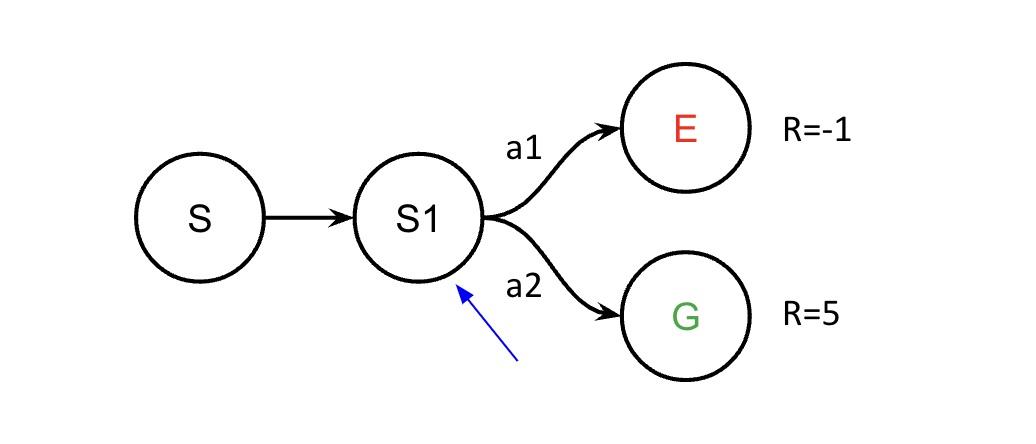}
    \label{fig:bor}
\end{figure}

\subsubsection{Fear}
The fear story is about an individual encountering an unfamiliar problem that could cause severe damage. The problem does not occur frequently, and this is reflected in state \( S_1 \), where there is only a 20\% chance that the agent will end up in the problem state \( P \). In the testing episode, the agent is designed to take the path \( S \xrightarrow{\text{{frwd}}} S_1 \xrightarrow{\text{{frwd}}} P \xrightarrow{\text{{frwd}}} E \). The appraisal analysis is conducted at the state \( P \).

\begin{figure}[H]
    \centering
    \includegraphics[width=8.5cm]{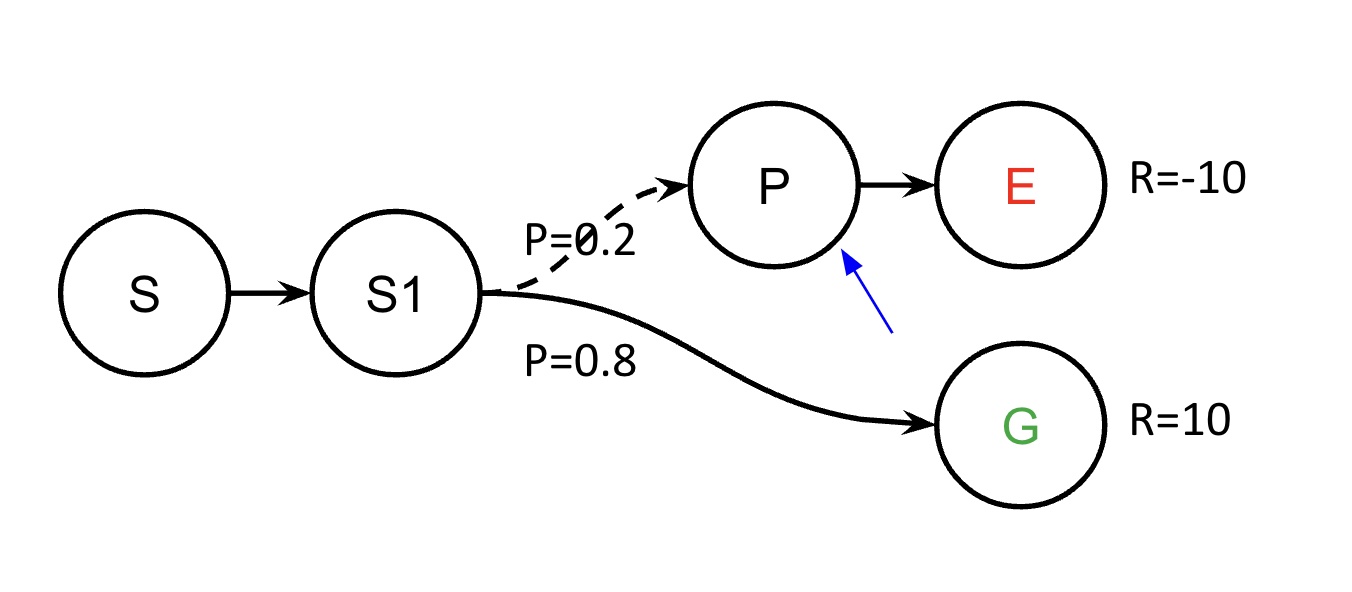}
    \label{fig:fear}
    \caption{MDP illustration for Fear}
\end{figure}

\subsubsection{Sadness}
This story is about a man who must say goodbye to a game that has kept him company for a long time. While he continues to be rewarded for playing the game, earning a reward of 10 and moving to state \( G \), his friends are now leaving the game. This is represented by transitioning to state \( P \) with a negative reward of -1. Ultimately, the game shuts down, and in the last 10 sessions of training, the reward at state \( E \) becomes -10. In the testing episode, the agent is designed to take the path \( S \xrightarrow{\text{{frwd}}} S_1 \xrightarrow{\text{{frwd}}} P \xrightarrow{\text{{frwd}}} E \). The appraisal analysis is conducted at the state \( P \).

\begin{figure}[H]
    \centering
    \includegraphics[width=8.5cm]{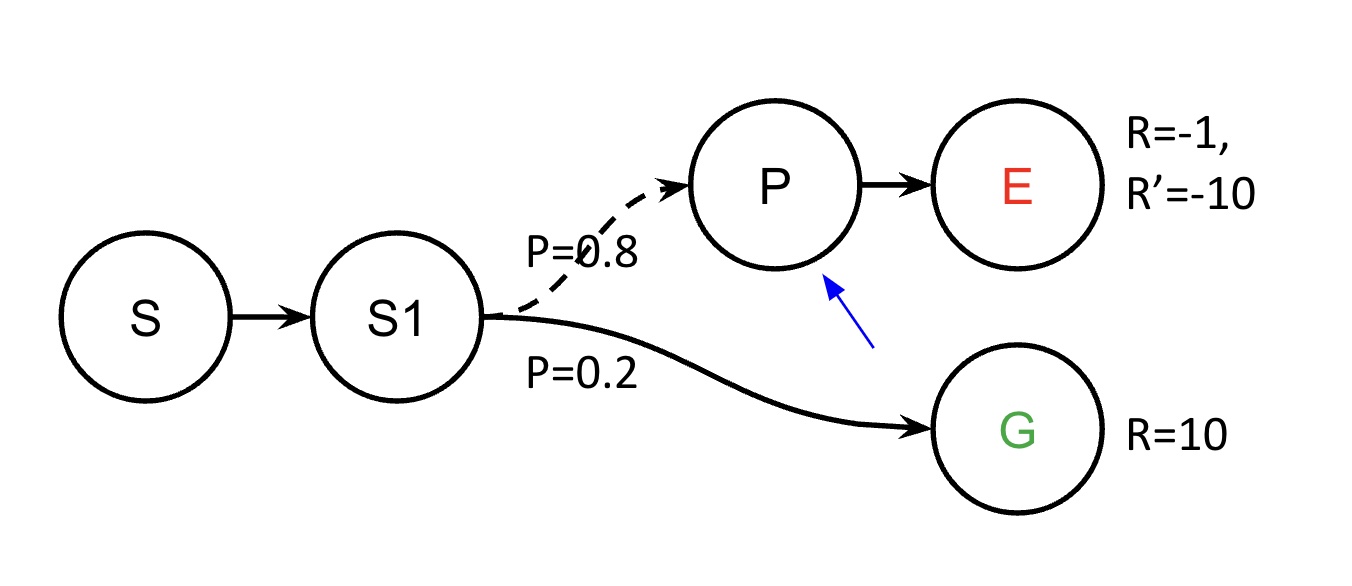}
    \label{fig:sad}
    \caption{MDP illustration for Sadness}
\end{figure}

\subsubsection{Shame}
The shame story is about an individual who takes an unusual action \( a_2 \) by opening a suspicious website. Typically, it is not immediately malicious, resulting in an 80\% chance of following the path \( S_2 \xrightarrow{\text{{frwd}}} G \). However, it can also lead to damage, represented by \( S_2 \xrightarrow{\text{{frwd}}} E \). The designed path for the agent in the testing episode is \( S \xrightarrow{\text{{frwd}}} S_1 \xrightarrow{a_2} S2 \xrightarrow{\text{{frwd}}} E \). The appraisal analysis is conducted at the state \( E \).

\begin{figure}[H]
    \centering
    \includegraphics[width=8.5cm]{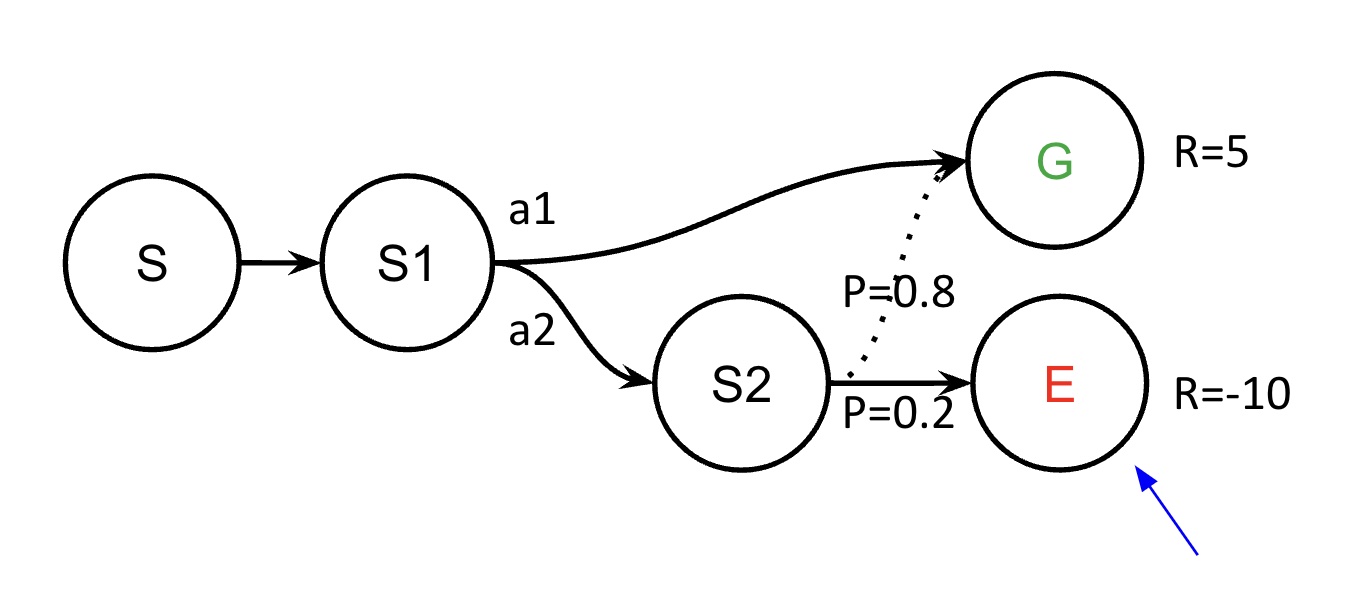}
    \label{fig:shame}
    \caption{MDP illustration for Shame}
\end{figure}

\subsubsection{Anxiety}
The anxiety story is about a freshman who encounters continuous system errors, and he doesn't know what to do. Represented by MDP, in the testing episode, the agent goes for \( S \xrightarrow{\text{{frwd}}} S_1 \xrightarrow{\text{{frwd}}} P \xrightarrow{\text{{frwd}}} E \).

\begin{figure}[H]
    \centering
    \includegraphics[width=8.5cm]{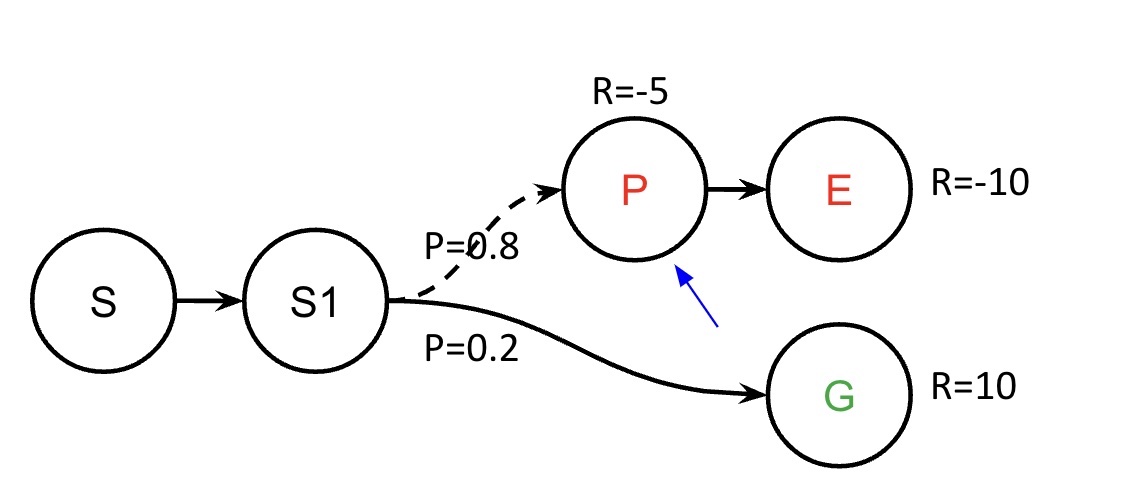}
    \label{fig:anxiety}
    \caption{MDP illustration for Anxiety}
\end{figure}

\subsubsection{Despair}
The despair story is about a freshman who got an unfamiliar system error and immediately failed the exam. The agent goes for \( S \xrightarrow{\text{{frwd}}} S_1 \xrightarrow{\text{{frwd}}} P \xrightarrow{\text{{frwd}}} E \) in the testing episode.

\begin{figure}[H]
    \centering
    \includegraphics[width=8.5cm]{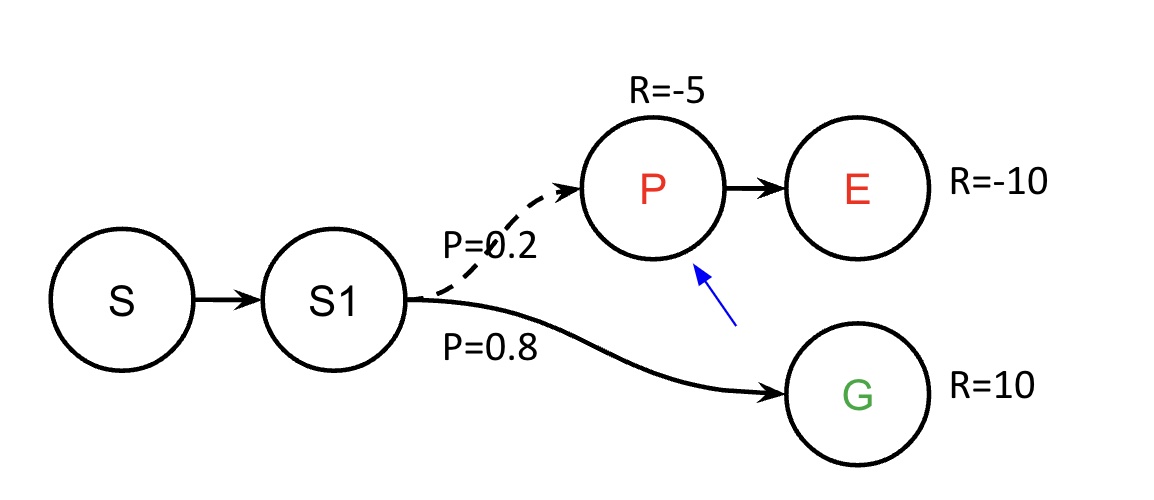}
    \label{fig:despair}
    \caption{MDP illustration for Despair}
\end{figure}

\subsubsection{Irritation}
The irritation story is about a senior student who is familiar with the school system. Thus, when the problem occurs, he knows what to do by choosing either \( a_1 \), which is to do nothing and fail the exam, or choosing \( a_2 \), which is to report the issue and get another chance for the exam. 

\begin{figure}[H]
    \centering
    \includegraphics[width=8.5cm]{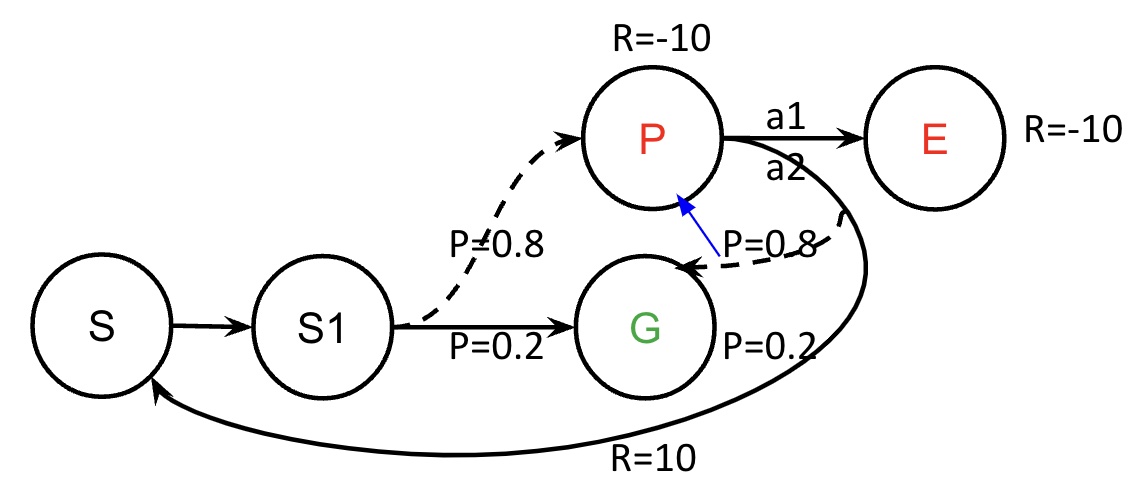}
    \label{fig:irritation}
    \caption{MDP illustration for Irritation}
\end{figure}

\subsubsection{Rage}

The rage story is also about a senior student who knows how to resolve the system problems. It is similar to irritation, he can choose \( a_1 \) or \( a_2 \) 

\begin{figure}[H]
    \centering
    \includegraphics[width=8.5cm]{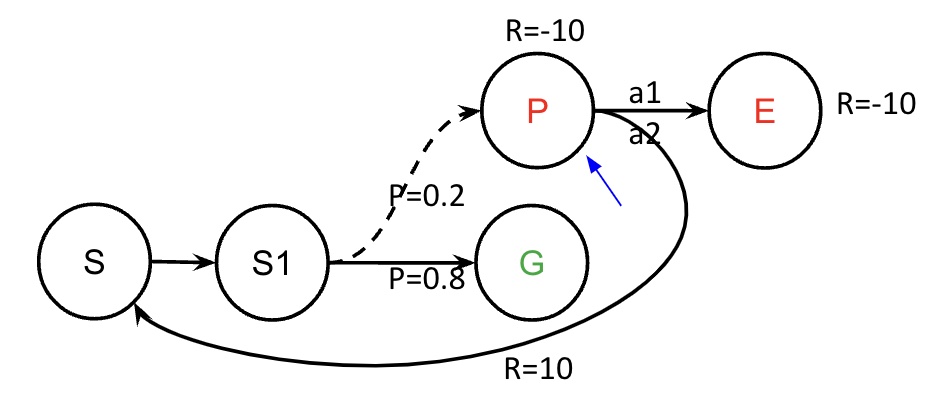}
    \label{fig:rage}
    \caption{MDP illustration for Rage}
\end{figure}

\subsection{Codes}
The codes associated with this paper are available on GitHub. These codes include implementations of the algorithms, experiments, and additional supporting material that can aid in understanding and reproducing the research. A comprehensive guide on how to use the codes can also be found within the repository. The codes can be accessed through the following \url{https://github.com/Eurus-J-Zhang/Appraisal_RL/}.

\end{document}